\documentclass[ALICE,manyauthors]{cernphprep}
\usepackage[comma,square,numbers,sort&compress]{natbib}

\usepackage{amsmath, amsthm, amssymb}
\usepackage[inline]{enumitem}
\usepackage{hyperref}
\usepackage{xspace}
\usepackage[T1]{fontenc}
\usepackage{epstopdf}
\begin{document}
%

\newcommand{\pp}           {pp\xspace}
\newcommand{\ppbar}        {\mbox{$\mathrm {p\overline{p}}$}\xspace}
\newcommand{\XeXe}         {\mbox{Xe--Xe}\xspace}
\newcommand{\PbPb}         {\mbox{Pb--Pb}\xspace}
\newcommand{\pA}           {\mbox{pA}\xspace}
\newcommand{\pPb}          {\mbox{p--Pb}\xspace}
\newcommand{\AuAu}         {\mbox{Au--Au}\xspace}
\newcommand{\dAu}          {\mbox{d--Au}\xspace}

\newcommand{\s}            {\ensuremath{\sqrt{s}}\xspace}
\newcommand{\snn}          {\ensuremath{\sqrt{s_{\mathrm{NN}}}}\xspace}
\newcommand{\pt}{p_{\mathrm{T}}}
\newcommand{\meanpt}       {\langle p_{\mathrm{T}}\rangle}
\newcommand{\ycms}         {\ensuremath{y_{\rm cms}}\xspace}
\newcommand{\ylab}         {\ensuremath{y_{\rm lab}}\xspace}
\newcommand{\etarange}[1]  {\mbox{$\left | \eta \right |~<~#1$}}
\newcommand{\yrange}[1]    {\mbox{$\left | y \right |~<~#1$}}
\newcommand{\etalab}{\eta_{\rm{lab}}}
\newcommand{\dndy}         {\ensuremath{\mathrm{d}N_\mathrm{ch}/\mathrm{d}y}\xspace}
\newcommand{\dndeta}       {\mathrm{d}N_\mathrm{ch}/\mathrm{d}\eta}
\newcommand{\avdndeta}     {\ensuremath{\langle\dndeta\rangle}\xspace}
\newcommand{\dNdy}         {\ensuremath{\mathrm{d}N_\mathrm{ch}/\mathrm{d}y}\xspace}
\newcommand{\Npart}        {\ensuremath{N_\mathrm{part}}\xspace}
\newcommand{\Ncoll}        {\ensuremath{N_\mathrm{coll}}\xspace}
\newcommand{\dEdx}         {\ensuremath{\textrm{d}E/\textrm{d}x}\xspace}
\newcommand{\RpPb}         {\ensuremath{R_{\rm pPb}}\xspace}

\newcommand{\nineH}        {$\sqrt{s}~=~0.9$~Te\kern-.1emV\xspace}
\newcommand{\seven}        {$\sqrt{s}~=~7$~Te\kern-.1emV\xspace}
\newcommand{\treize}        {$\sqrt{s}~=~13$~Te\kern-.1emV\xspace}
\newcommand{\twoH}         {$\sqrt{s}~=~0.2$~Te\kern-.1emV\xspace}
\newcommand{\twosevensix}  {$\sqrt{s}~=~2.76$~Te\kern-.1emV\xspace}
\newcommand{\five}         {$\sqrt{s}~=~5.02$~Te\kern-.1emV\xspace}
\newcommand{\twosevensixnn}{$\sqrt{s_{\mathrm{NN}}}~=~2.76$~Te\kern-.1emV\xspace}
\newcommand{\fivenn}       {$\sqrt{s_{\mathrm{NN}}}~=~5.02$~Te\kern-.1emV\xspace}
\newcommand{\eightnn}       {$\sqrt{s_{\mathrm{NN}}}~=~8.16$~Te\kern-.1emV\xspace}
\newcommand{\eightnnNS}       {$\sqrt{s_{\mathrm{NN}}}~=~8.16$~Te\kern-.1emV\xspace}
\newcommand{\LT}           {L{\'e}vy-Tsallis\xspace}
\newcommand{\GeVc}         {Ge\kern-.1emV/$c$\xspace}
\newcommand{\MeVc}         {Me\kern-.1emV/$c$\xspace}
\newcommand{\TeV}          {Te\kern-.1emV\xspace}
\newcommand{\GeV}          {Ge\kern-.1emV\xspace}
\newcommand{\MeV}          {Me\kern-.1emV\xspace}
\newcommand{\GeVmass}      {Ge\kern-.2emV/$c^2$\xspace}
\newcommand{\MeVmass}      {Me\kern-.2emV/$c^2$\xspace}
\newcommand{\lumi}         {\ensuremath{\mathcal{L}}\xspace}

\newcommand{\ITS}          {\rm{ITS}\xspace}
\newcommand{\TOF}          {\rm{TOF}\xspace}
\newcommand{\ZDC}          {\rm{ZDC}\xspace}
\newcommand{\ZDCs}         {\rm{ZDCs}\xspace}
\newcommand{\ZNA}          {\rm{ZNA}\xspace}
\newcommand{\ZNC}          {\rm{ZNC}\xspace}
\newcommand{\SPD}          {\rm{SPD}\xspace}
\newcommand{\SDD}          {\rm{SDD}\xspace}
\newcommand{\SSD}          {\rm{SSD}\xspace}
\newcommand{\TPC}          {\rm{TPC}\xspace}
\newcommand{\TRD}          {\rm{TRD}\xspace}
\newcommand{\VZERO}        {\rm{V0}\xspace}
\newcommand{\VZEROA}       {\rm{V0A}\xspace}
\newcommand{\VZEROC}       {\rm{V0C}\xspace}
\newcommand{\Vdecay} 	   {\ensuremath{V^{0}}\xspace}

\newcommand{\ee}           {\ensuremath{e^{+}e^{-}}} 
\newcommand{\pip}          {\ensuremath{\pi^{+}}\xspace}
\newcommand{\pim}          {\ensuremath{\pi^{-}}\xspace}
\newcommand{\kap}          {\ensuremath{\rm{K}^{+}}\xspace}
\newcommand{\kam}          {\ensuremath{\rm{K}^{-}}\xspace}
\newcommand{\pbar}         {\ensuremath{\rm\overline{p}}\xspace}
\newcommand{\kzero}        {\ensuremath{{\rm K}^{0}_{\rm{S}}}\xspace}
\newcommand{\lmb}          {\ensuremath{\Lambda}\xspace}
\newcommand{\almb}         {\ensuremath{\overline{\Lambda}}\xspace}
\newcommand{\Om}           {\ensuremath{\Omega^-}\xspace}
\newcommand{\Mo}           {\ensuremath{\overline{\Omega}^+}\xspace}
\newcommand{\X}            {\ensuremath{\Xi^-}\xspace}
\newcommand{\Ix}           {\ensuremath{\overline{\Xi}^+}\xspace}
\newcommand{\Xis}          {\ensuremath{\Xi^{\pm}}\xspace}
\newcommand{\Oms}          {\ensuremath{\Omega^{\pm}}\xspace}
\newcommand{\degree}       {\ensuremath{^{\rm o}}\xspace}

\newcommand{\Jpsi}{\rm{J}/\psi}
\newcommand{\psit}{\psi(2{\rm S})}
\newcommand{\psip}{\psi(2{\rm S})}
\newcommand{\Tm}{\ensuremath{\rm{T}\times}\rm{m}\xspace}
\newcommand{\abs}[1]{\left\vert #1 \right\vert}
\newcommand{\ave}[1]{\langle #1 \rangle}
\newcommand{\zv}{z_{\rm vtx}}
\newcommand{\Ntr}{N_{\rm tracklet}}
\newcommand{\Ntrc}{N_{\rm tracklet}^{\rm corr}}
\newcommand{\Ae}{A\varepsilon}
\newcommand{\mumu}{\mu^+\mu^-}
\newcommand{\BR}{\rm{BR}_{\Jpsi\rightarrow\mumu}}
\newcommand{\Fnorm}{F_{\rm{norm}}}

\begin{titlepage}
\PHyear{2020}       
\PHnumber{058}      
\PHdate{15 April}  

\title{$\Jpsi$ production as a function of charged-particle multiplicity in \pPb collisions at $\mathbf{\sqrt{\textit{s}_{\mathrm{\textbf{NN}}}}~=~8.16}$~Te\kern-.1emV}
\ShortTitle{Multiplicity dependent $\Jpsi$ production in \pPb at \eightnnNS}   

\Collaboration{ALICE Collaboration\thanks{See Appendix~\ref{app:collab} for the list of collaboration members}}
\ShortAuthor{ALICE Collaboration} 

\begin{abstract}
Inclusive $\Jpsi$ yields and average transverse momenta in \pPb collisions at a center-of-mass energy per nucleon pair \eightnnNS~are measured as a function of the charged-particle pseudorapidity density with ALICE. The $\Jpsi$ mesons are reconstructed at forward ($2.03 < \ycms < 3.53$) and backward ($-4.46 < \ycms < -2.96$) center-of-mass rapidity in their dimuon decay channel while the charged-particle pseudorapidity density is measured around midrapidity.
The $\Jpsi$ yields at forward and backward rapidity normalized to their respective average values increase with the  normalized charged-particle pseudorapidity density, the former showing a weaker increase than the latter. The normalized average transverse momenta at forward and backward rapidity manifest a steady increase from low to high charged-particle pseudorapidity density with a saturation beyond the average value.
\end{abstract}
\end{titlepage}

\setcounter{page}{2} 


\section{Introduction} 
\label{sec:introduction}
Quarkonium states have long been considered as probes of the Quark--Gluon Plasma (QGP) produced in ultra-relativistic heavy-ion collisions~\cite{BraunMunzinger:2007zz}. 
The large color-charge density in the plasma prevents the formation of bound states, 
in an analogous process to the Debye screening for electromagnetic processes~\cite{Matsui:1986dk}. 
The suppression of $\Jpsi$ production in nucleus--nucleus (AA) with respect to proton--proton (\pp) collisions was observed by several experiments~\cite{Alessandro:2004ap, Arnaldi:2007zz, Adare:2011yf, Abelev:2012rv, Abelev:2013ila, Adamczyk:2013tvk, Chatrchyan:2012np, Aad:2010aa, Adam:2016rdg}. 
To determine whether the origin of this suppression is the influence of the QGP or of Cold Nuclear Matter (CNM), data on proton(deuteron)--nucleus 
collisions are also scrutinized. 

The measurements in \pPb collisions at the LHC show a suppression of $\Jpsi$ production~\cite{Abelev:2013yxa,Acharya:2018kxc,Aaij:2013zxa}, with respect to \pp collisions, at low transverse momentum ($\pt$) and forward center-of-mass rapidity (p-going direction, positive $\ycms$), consistent with various combinations of CNM effects: 
 modification of the parton distribution functions (PDFs) in nuclei, i.e.\ shadowing~\cite{Albacete:2016veq,Kusina:2017gkz}, the Color-Glass Condensate (CGC)~\cite{Ma:2017rsu,Ducloue:2016ywt}, or coherent parton energy loss~\cite{Arleo:2014oha}. 
The measurement of $\psit$ production in \pPb collisions~\cite{Abelev:2014zpa} exhibits a larger suppression, with respect to \pp collisions, than the one measured for $\Jpsi$, both at forward and backward rapidity, which was not expected from CNM predictions.
This effect is reproduced by models which consider the break-up of the bound quark--anti-quark pair via interactions with the final-state comoving particles~\cite{Ferreiro:2014bia,Chen:2016dke}.  

The \pPb data at the center-of-mass energy per nucleon--nucleon collision of $\snn = 5.02$~TeV \cite{Adam:2015jsa,Adam:2016ohd} showed that these effects depend on the centrality of the collision, as estimated from the energy deposited in the Zero Degree Calorimeter in the Pb-going direction~\cite{Adam:2014qja}, and/or the produced charged-particle multiplicity~\cite{Adamova:2017uhu}. 
An increase of the normalized $\Jpsi$ and $\Upsilon$~\cite{Adamova:2017uhu,Chatrchyan:2013nza,Aaboud:2017cif} yields, to their respective average values, with the normalized charged-particle multiplicity is observed, similarly to the results from \pp collisions~\cite{Abelev:2012rz,Chatrchyan:2013nza,Aaboud:2017cif}. 
%
The increase of the $\Jpsi$ (prompt and non-prompt) normalized yields was observed to be similar to the increase for D mesons~\cite{Adam:2015ota,Adam:2016mkz}, suggesting that the origin of the trend is common for charm and beauty production, and that hadronization does not play a dominant influence on this measurement.
The excited-to-ground state ratios, $\Upsilon{\rm (nS)}/\Upsilon{\rm (1S)}$, were found to decrease with increasing charged-particle multiplicity, which was not expected from CNM predictions~\cite{Chatrchyan:2013nza,Aaboud:2017cif}. 

The measurements of two-particle angular correlations in small systems have shown interesting structures in the angular correlation function. 
A near-side ridge, located at $(\Delta \varphi) \approx 0$, is observed in high-multiplicity \pp~\cite{Khachatryan:2010gv} and \pPb~\cite{CMS:2012qk} collisions, accompanied by an away-side structure, located at $\Delta \varphi \approx \pi$ and exceeding the away-side jet contribution, in \pPb collisions~\cite{Abelev:2012ola,Aad:2012gla}. These structures are reminiscent of those in \PbPb data~\cite{Aamodt:2011by}, interpreted as signatures of the collective motion of the particles during the hydrodynamic evolution of the hot and dense medium. 
Correlations of $\Jpsi$ (at large rapidity) and charged particles (at midrapidity) in \pPb collisions~\cite{Acharya:2017tfn, Sirunyan:2018kiz} revealed persisting long-range correlation structures at high $\pt$, similar to those observed with charged hadrons. The corresponding elliptic flow coefficients are found to be positive and of comparable magnitude to those measured in \PbPb collisions~\cite{Acharya:2017tgv,Acharya:2018pjd,Aaboud:2018ttm}, indicating that the mechanism at its origin could be similar in both collision systems.

This letter reports the measurement of the multiplicity-differential inclusive $\Jpsi$ yield and average transverse momentum in \pPb collisions at $\snn=8.16$~TeV. 
The $\Jpsi$ mesons are reconstructed at forward and backward center-of-mass rapidities in their dimuon decay channel. The charged-particle pseudorapidity density is measured around midrapidity.
It complements and extends previous $\Jpsi$ measurements performed as a function of the collision centrality and the charged-particle multiplicity at $\snn=5.02$~TeV~\cite{Adamova:2017uhu,Adam:2015jsa}. 
The classification of events as a function of their charged-particle pseudorapidity density enables the scrutiny of rare events, corresponding to the 0.01--0.04\% highest multiplicities in the collision. 
This allows \pPb~events to be studied from low multiplicities, similar to those of pp collisions, up to very large multiplicities corresponding to  $\sim 100$ produced charged particles per rapidity unit, similar to those of peripheral \PbPb~collisions, which exhibit collective-like effects.

\section{Experimental setup and data samples}
\label{sec:expData}
In this section, the detector subsystems relevant for this analysis are presented. 
A complete description of the ALICE detector and its performance  can be found in~\cite{Aamodt:2008zz,Abelev:2014ffa}. 

The muon spectrometer~\cite{Aamodt:2008zz,Abelev:2014ffa} covers the pseudorapidity window of $-4.0 < \eta < -2.5$ and consists of: a 4~m long composite front absorber, corresponding to about 10 interaction lengths ($10 \ \lambda_{\rm int}$), starting at 90~cm from the nominal interaction point,  ten layers of muon tracking chambers (MCH), coupled to a dipole magnet with a 3 Tm field integral, and four layers of muon trigger chambers (MTR).
The MCH and MTR systems are separated by an additional iron wall of about $7.2 \ \lambda_{\rm int}$ that absorbs the remaining hadronic and low-momentum particle contamination. A rear absorber positioned downstream of the MTR filters out the background from beam-gas interactions. 
A conical absorber surrounds the beam pipe and protects the spectrometer against secondary particles produced mainly by large-$\eta$ primary particles interacting with the beam pipe.

The Silicon Pixel Detector (SPD)~\cite{Aamodt:2010aa} is the innermost part of the Inner Tracking System (ITS). It consists of two cylindrical silicon pixel layers at radial distances of $3.9 \mbox{ and } 7.6 \textup{ cm}$ from the beam line. The respective pseudorapidity coverage of the two layers are $\abs{\eta}<2$ and $\abs{\eta}<1.4$.  The SPD is used to reconstruct the primary vertex and to measure the charged-particle pseudorapidity density at midrapidity. 

The V0 scintillator arrays~\cite{Abbas:2013taa} are located at each side of the interaction point, covering the pseudorapidity ranges of $-3.7 < \eta < -1.7$ and $2.8 < \eta < 5.1$. In this analysis, the V0 provides an online trigger and helps to reject contamination from beam-gas events.

The neutron Zero Degree Calorimeter (ZDC)~\cite{Aamodt:2008zz}  located at about 112.5 m on either side from the interaction point are used to reject electromagnetic interactions and beam-induced background.


The results presented in this letter are obtained with data recorded during the \pPb run at \eightnnNS in 2016. The $\Jpsi$ are reconstructed in the dimuon channel with data taken in two different beam configurations. 
Due to the asymmetry of the beam energy per nucleon in \pPb collisions at the LHC, the nucleon--nucleon center-of-mass rapidity frame is shifted by $\Delta y = 0.465$ in the direction of the proton beam. As a consequence, the $\Jpsi$ are measured in the forward rapidity range of $2.03 < \ycms < 3.53$ (with protons going in the direction of the muon spectrometer, p-going direction) and in the backward rapidity region  $-4.46 < \ycms < -2.96$ (Pb-going direction).
Events used in this analysis were collected with a dedicated dimuon trigger which requires the coincidence of 
signals in both V0 arrays (minimum bias trigger, MB) with at least two opposite-sign muons registered in the MTR. 
The trigger has an adjustable online threshold, which for this data sample was set to only accept muons with transverse momenta $\pt > 0.5$~GeV/$c$ ($\pt$ for which an efficiency of 50\% is reached). The $\pt$ differential single-muon trigger efficiency reaches a plateau of $\sim 96\%$ at $\pt \sim 1.5$~GeV/$c$.
In this data-taking period, the maximum pile-up probability was about 4\%. A dedicated event-selection strategy---exploiting the signals in the V0 and the ZDC, the correlation of the number of clusters and track segments reconstructed in the SPD, as well as an algorithm to tag events with multiple vertices---allowed us to keep the pile-up below 0.5\% for the analysed events, even at large multiplicities. 
The data sample analyzed corresponds to an integrated luminosity of $\mathcal{L}_{\rm int}= 7.2 \pm 0.2 \text{ nb}^{-1}$ ($10.6 \pm 0.3\text{ nb}^{-1}$) for the p-going (Pb-going) configuration~\cite{ALICE-PUBLIC-2018-002}.

\section{Charged-particle multiplicity measurement}
\label{sec:multiplicity}
The charged-particle pseudorapidity density ($\dndeta$) is measured at midrapidity exploiting the information provided by the SPD detector~\cite{ALICE:2012xs,Acharya:2018egz}. It is evaluated by counting the number of tracklets ($\Ntr$), i.e. track segments joining pairs of hits in the two layers of the SPD pointing to the primary vertex. The primary vertex is also computed with the SPD information. 
To minimize non-uniformities in the SPD acceptance, only events with a $z$-vertex position determined within $|\zv|<10$~cm are considered, and tracklets are counted within $|\eta|<1$. 

The raw $\Ntr$ counts are corrected ($\Ntrc$) for the variation of the detector conditions with time (fraction of active SPD channels) and its limited acceptance as a function of $\zv$ using a data-driven event-by-event correction~\cite{Abelev:2012rz,Adam:2015ota}. This correction ensures a uniform response as a function of $\zv$. 
In this analysis, the correction is done by renormalising the $\Ntr(\zv)$  distributions to the overall maximum with a Poissonian smearing to account for the fluctuations. 
The events are sliced in $\Ntrc$ intervals. 
Monte Carlo (MC) simulations using the DPMJET~\cite{Roesler:2000he} event generator and the GEANT3 transport code~\cite{Brun:1082634} are used to estimate $\dndeta$ from $\Ntrc$. 
A second order polynomial correlation is assumed between these two quantities for the full $\Ntrc$ interval. 
Several sources of systematic uncertainty were taken into account.
Possible deviations from the second order polynomial correlation were estimated by using other functions to quantify the correlation or MC averages in each interval, with values ranging from 0.1\% at intermediate multiplicities to 6.9\% (5.8\%) at the lowest (highest) multiplicity intervals.
The systematic uncertainty on the residual $\zv$ dependence due to differences between data and MC amounts to 3\%.
Finally, the event generator influence was considered and evaluated by comparing the DPMJET simulations with events generated in EPOS ~\cite{Pierog:2013ria}, resulting in a 2\%  uncertainty.

The average charged-particle pseudorapidity density, $\avdndeta$, in non-single diffractive (NSD) events was obtained from an independent analysis and amounts to $\avdndeta=20.33 \pm 0.83$ ($20.32 \pm 0.83$) in p--Pb (Pb--p) collisions for $|\eta|<1$~\cite{Acharya:2018egz}, where the quoted uncertainty is systematic. 

Table~\ref{tab:unc_mult} summarizes the contributions to the normalized charged-particle multiplicity uncertainty. 
The total uncertainty is evaluated assuming that the different sources are uncorrelated. 
\begin{table}[!htbp]
   \centering
      \caption{Sources of systematic uncertainties on the normalized charged-particle multiplicity. 
      For the $\Ntrc$ to $\dndeta$ correlation an interval is quoted, varying with multiplicity, with a different maximum uncertainty for the Pb(p)-going configuration. 
	   \label{tab:unc_mult}
	   }
	   \begin{tabular}{@{} c | c @{}} 
	\hline
	Source  				& $|\eta| < 1$   \\ \hline
	$\Ntrc$ to $\dndeta$	correlation &	0.1 -- 6.9(5.8)\% 	\\
	$z$-vertex dependence & 3\% \\
	Monte Carlo event generator & 2\% \\ 
	$\avdndeta$ & 4\% \\
						 \hline
   \end{tabular}
\end{table}

\section{$\Jpsi$ measurement}
\label{sec:analysis}
The normalized $\Jpsi$ yield, i.e.\ the yield in each multiplicity interval $i$ normalized to the multiplicity-integrated value, is evaluated as
\begin{equation}
 \frac { {\rm d}N^i/{\rm d}y }{ \langle {\rm d}N/{\rm d}y \rangle } = 
 \frac{ N^i_{\Jpsi} }{ N_{\Jpsi} } 
 \frac{N_{\rm{MB}}^{\rm{eq}}}{N^{i\rm{, eq}}_{\rm{MB}}} 
 \frac{(A\varepsilon)_{\Jpsi}}{(A\varepsilon)^i_{\Jpsi}}
 \frac{\varepsilon_{\rm{MB}}^i}{\varepsilon_{\rm{MB}}} \, ,
\label{eq:relYield}
\end{equation}
from the reconstructed number of $\Jpsi$, $N_{\Jpsi}$, 
the number of minimum bias (MB) events equivalent to the analysed dimuon sample, $N^{\rm{eq}}_{\rm{MB}}$, the $\Jpsi$ acceptance and efficiency correction, $(A\varepsilon)_{\Jpsi}$, and the NSD event selection efficiency in the minimum bias sample, $\varepsilon_{\rm{MB}}$. 

The $\Jpsi$ are reconstructed for each multiplicity interval by combining opposite-sign muons and computing the invariant mass of the pairs. 
The muon identification is ensured by requiring that the track candidates reconstructed in the MCH have a matching track segment in the MTR. 
Furthermore, the individual tracks must fulfill the following criteria to make sure they are within the acceptance of the spectrometer: their radial distance from the beam axis at the end of the front absorber is within $17.6 < R_{\rm{abs}} < 89.5 \textup{ cm}$ and their pseudorapidity in the detector reference frame is within $-4 < \eta < -2.5$.

\begin{figure}[!htbp]
   \centering
 \includegraphics[width=0.485\textwidth]{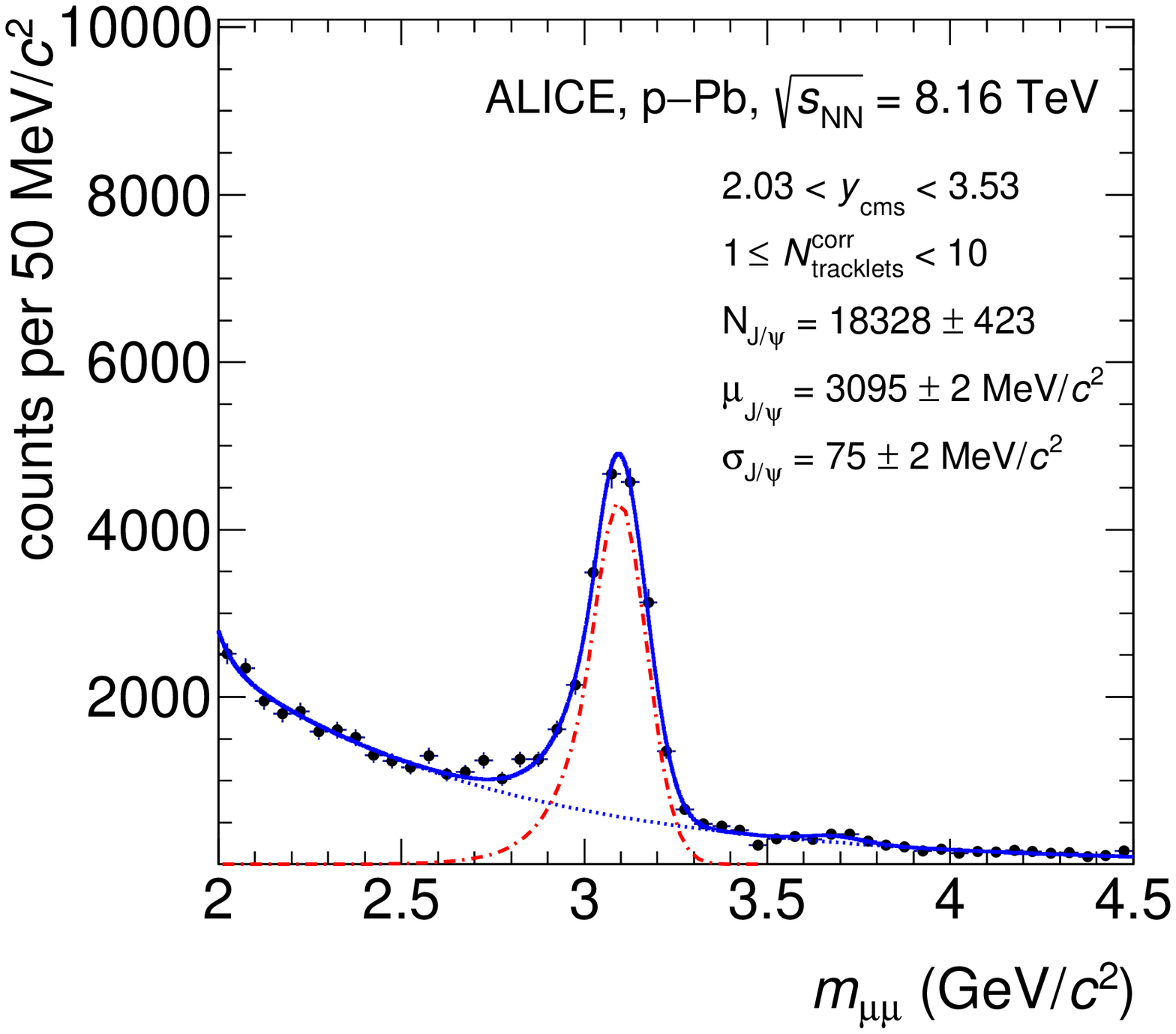}
   \includegraphics[width=0.485\textwidth]{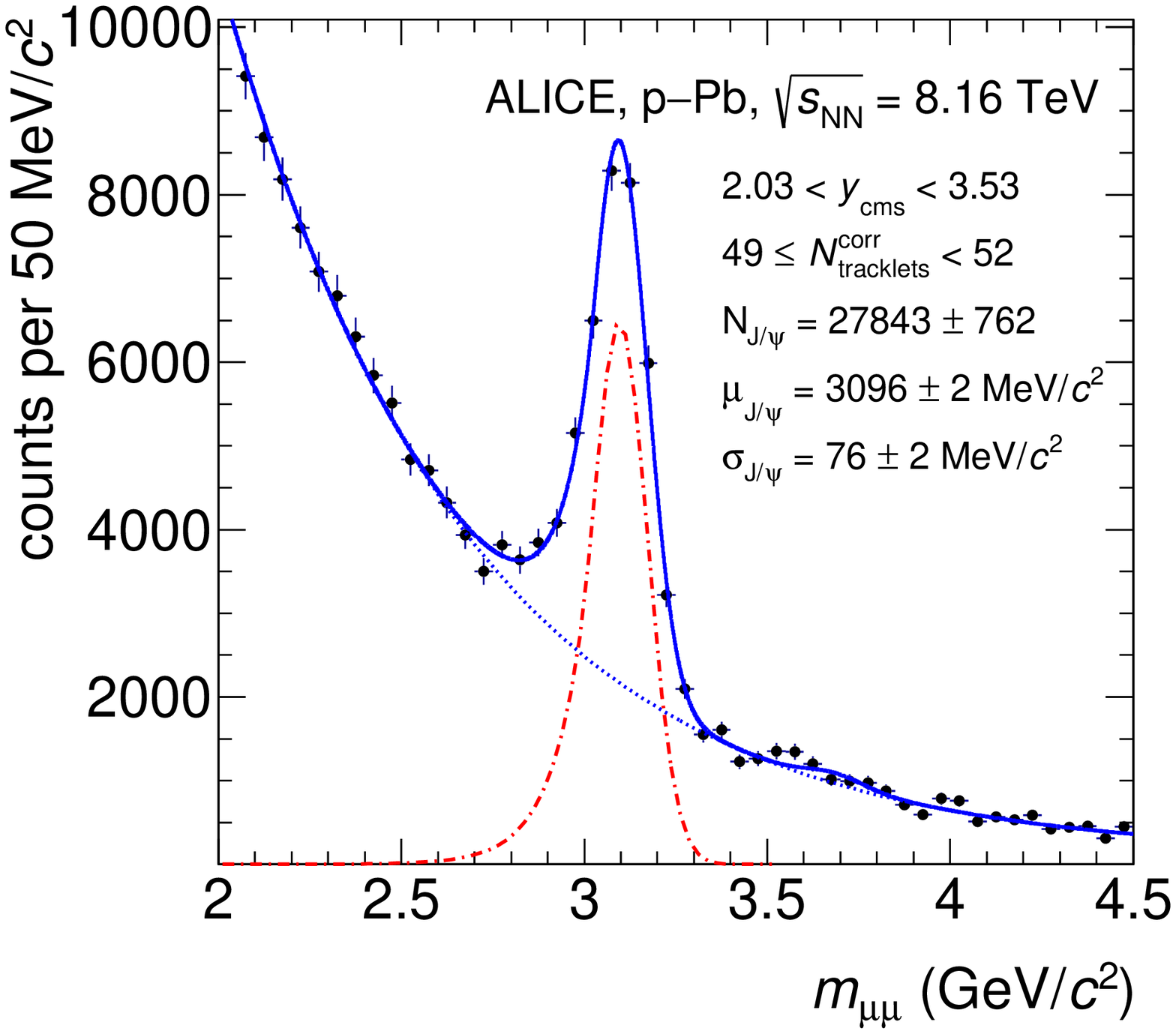}\\
   \includegraphics[width=0.485\textwidth]{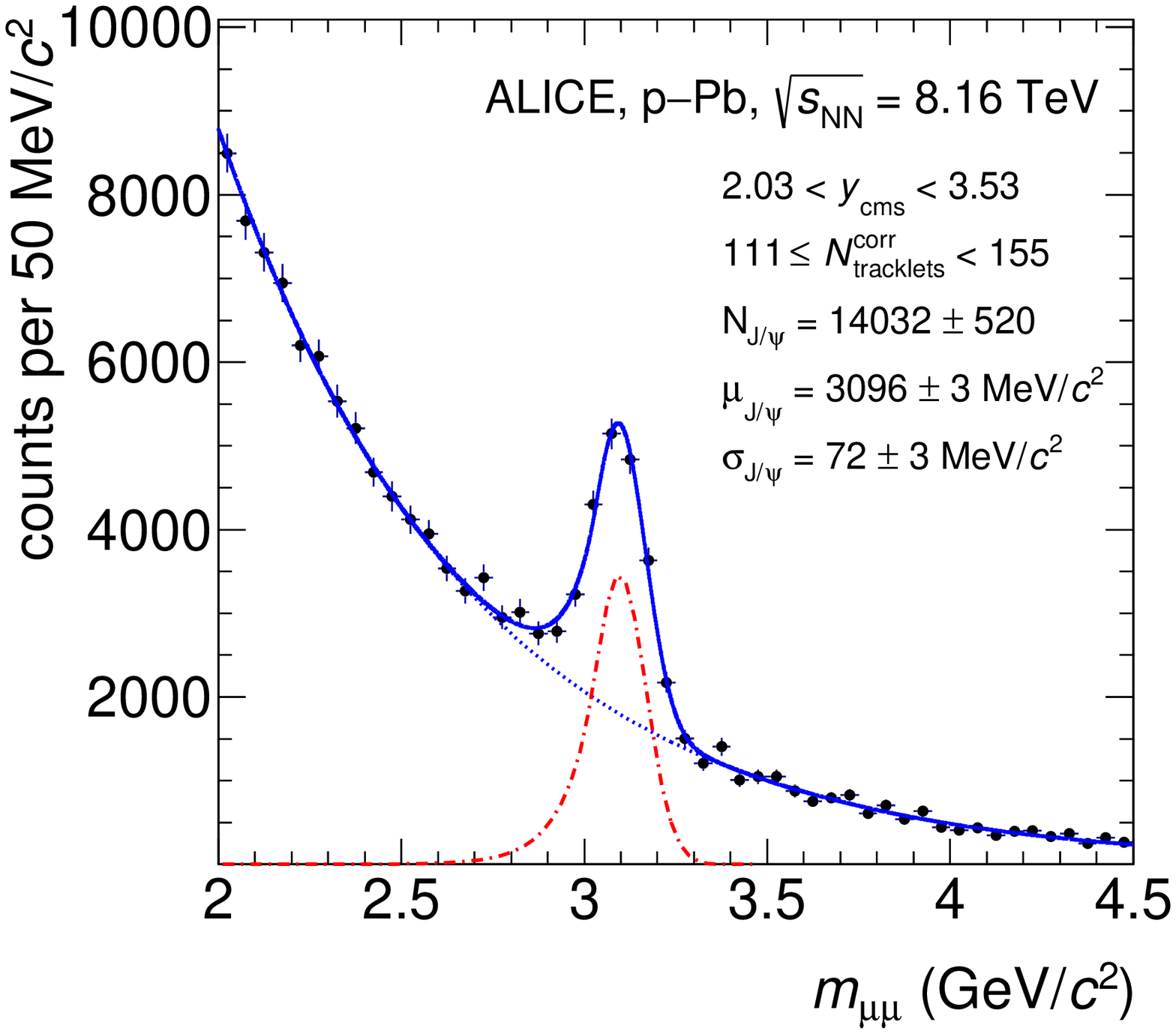}
   \caption{
      \label{fig:JpsiFits_multBins}
      Opposite-sign muon pair invariant mass distributions for selected multiplicity intervals, corrected for the $\Jpsi$ acceptance and efficiency, at forward rapidity. 
      The distributions are shown together with a typical fit function (solid line, see text for details). The $\Jpsi$ signal contribution is also depicted by a dot-dashed red line, and the background by a dotted line. 
      }
\end{figure}

To extract the signal, the invariant-mass distributions are first corrected for the $\Jpsi$ acceptance times efficiency ($\Ae$), differentially in $\pt$ and $y$. The resulting distributions are then fitted with a superposition of $\Jpsi$ and $\psip$ signals and a background lineshape. 
Various combinations of lineshapes are used in order to evaluate the signal counts and their uncertainties. The two charmonium resonances are parametrized by a sum of either two Crystal Ball or two pseudo-Gaussian functions with power-law tails~\cite{ALICE-PUBLIC-2015-006}. 
The tail parametrizations are fixed to the values determined from either fits of the $\Jpsi$ signal from MC simulations or to values taken from fits to the multiplicity-integrated distribution in \pPb\ data at \eightnn~\cite{Acharya:2018kxc} and in pp data at \treize ~\cite{Acharya:2017hjh}. 
The tails obtained from fitting the multiplicity-integrated distributions using the Crystal Ball function are also considered, and fixed in the binned fits.
The $\Jpsi$ peak mean position and width are left free in the multiplicity-integrated fit, whilst the $\psip$ ones are bound to those of the $\Jpsi$ following the same procedure as in~\cite{Adam:2015rta}. 
Note that the $\psip$ yields obtained are not physical values, as the invariant-mass spectrum is corrected by the $\Ae$ correction for the $\Jpsi$. 
In the multiplicity-differential fits, the mass and width of the $\Jpsi$ peak are fixed to the integrated values to ensure the convergence of the fits in the few cases where statistical significance is low.
The background is parameterized by either a sum of two exponentials or the product of an exponential and a fourth-order polynomial. 
Two fit mass ranges are taken into account when computing the average number of $\Jpsi$ and its uncertainty: $ 1.7 <  m_{\mu\mu} < 4.8 \text{ GeV}/c^2 $ and $ 2.0 < m_{\mu\mu} < 5.0 \text{ GeV}/c^2 $. 
Examples of fits at low, intermediate, and high multiplicity for data in the rapidity range $2.03 < \ycms < 3.53$  are shown in Fig.~\ref{fig:JpsiFits_multBins}. 
The signal lineshape is found to be independent of multiplicity, while the background does change with multiplicity. Therefore, in order to minimize the uncertainty on the signal extraction, the same signal lineshape  is used in the fit function for both the numerator and denominator in Eq.~\ref{eq:relYield}.

The number of equivalent MB events $N^{\rm{eq}}_{\rm{MB}}$ is computed from the number of dimuon triggered events, $N_{\mu\mu}$, and the normalization factor of dimuon triggered to MB events (calculated as explained in next section) as $N^{\rm{eq}}_{\rm{MB}} = \Fnorm \cdot N_{\mu\mu}$. The number needs to be corrected for by the NSD event selection efficiency, $\varepsilon_{\rm{MB}}=(97 \pm 1) \%$~\cite{Acharya:2018egz}, to take into account the fraction of events without a reconstructed SPD vertex that are rejected. 
This factor $\varepsilon_{\rm{MB}}$ is found to be independent of the charged-particle multiplicity in all the intervals studied, with the exception of the lowest multiplicity interval, where it decreases by 1\%.

The $\Jpsi$ acceptance and efficiency correction is obtained from MC simulations as a function of $\pt$ and $\ycms$. 
The $\Jpsi$ are generated using $\pt$ and $\ycms$ distributions tuned to data~\cite{Acharya:2018kxc}. They are simulated to decay into a muon pair using EvtGen~\cite{Lange:2001uf}. The final state radiation is described with PHOTOS~\cite{Barberio:1993qi}. 
The acceptance and efficiency correction is independent of multiplicity in the measurement intervals. Therefore, when estimating the uncertainty on the MC input, only the possible variation of the input $\pt$ and $\ycms$ distributions is taken into account by using as input a subsample of the lower/higher multiplicity events.

\begin{figure}[!htbp]
   \centering
   \includegraphics[width=0.4\textwidth]{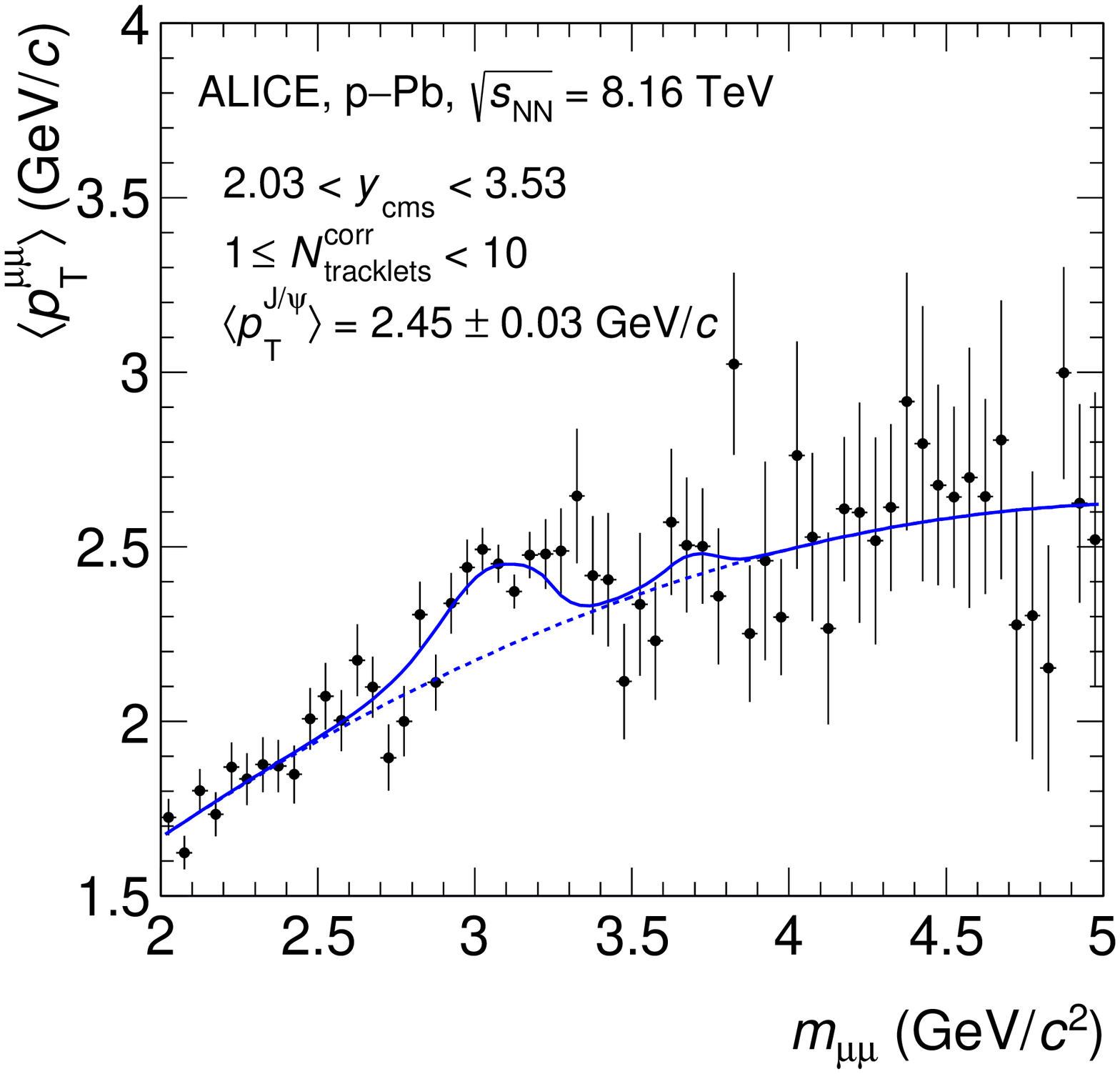}
   \includegraphics[width=0.4\textwidth]{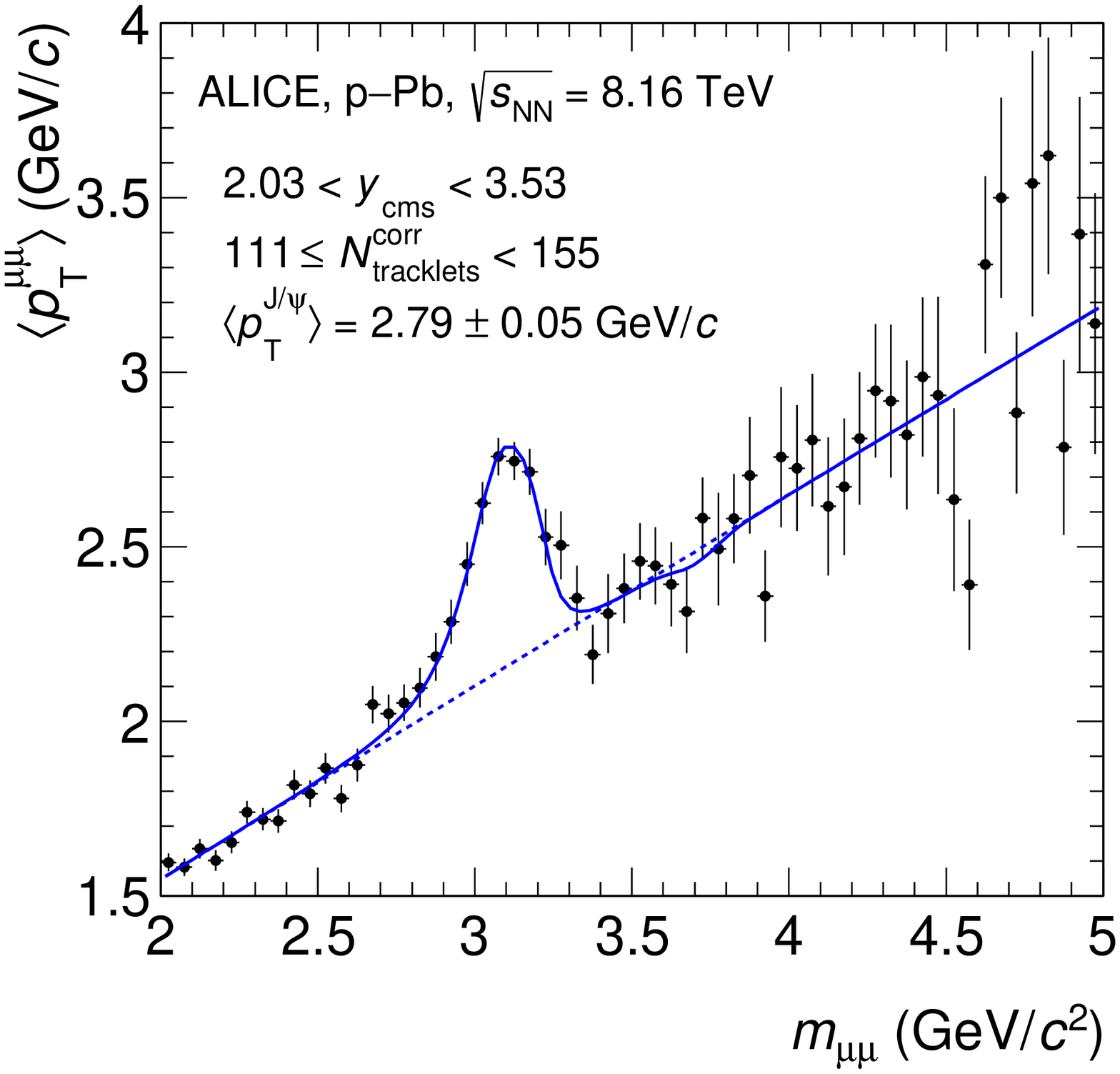}
   \caption{
      \label{fig:JpsiPtFits_multBins}
      Average transverse momentum of opposite-sign muon pairs for selected multiplicity intervals, corrected for the $\Jpsi$ acceptance and efficiency. 
      The distributions are shown together with a typical fit function (solid line, see text for details).
      }
\end{figure}

To extract the $\Jpsi$ mean transverse momentum $\ave{\pt^{\Jpsi}}$, the $\Ae$-corrected transverse momentum of the dimuon pair is fitted with the following function~\cite{Adamova:2017uhu}:
\begin{equation}
   \begin{array}{c l}
      \ave{\pt^{\mu\mu}}(m_{\mu\mu}) 	&= \alpha^{\Jpsi}(m_{\mu\mu}) \, \ave{\pt^{\Jpsi}} \\
				&+ \alpha^{\psi'}(m_{\mu\mu})\, \ave{\pt^{\psi'}} \\
				&+ \left(1-  \alpha^{\Jpsi}(m_{\mu\mu}) - \alpha^{\psi'}(m_{\mu\mu}) \right) \,\ave{\pt^{\rm bkgd}}(m_{\mu\mu}),\\  
 \end{array}
   \label{eq:meanPtFit}
\end{equation}
where the ratios of signal over the sum of signal and background of the two charmonium states $\alpha^{\Jpsi} = S^{\Jpsi}/(S^{\Jpsi} + S^{\psi'} + B)$ and $\alpha^{\psi'} = S^{\psi'}/(S^{\Jpsi} + S^{\psi'} + B)$ are fixed to the value extracted from fitting the invariant-mass spectrum corrected by the $\Jpsi$ $\Ae$. 
The background is described by a function $\ave{\pt^{\rm bkgd}}(m_{\mu\mu})$. 
Two functional forms are used: either a sum of two exponentials or the product of an exponential and a fourth-order polynomial. 
Note that  the $\ave{\pt^{\psi'}}$ does not represent a physical mean transverse momentum of the $\psip$ as the spectra are corrected by the $\Ae$ for $\Jpsi$. 
Figure~\ref{fig:JpsiPtFits_multBins} illustrates typical $ \ave{\pt^{\mu\mu}}$ distributions for selected multiplicity intervals.

\section{Systematic uncertainties}
\label{sec:uncertainties}

The following sources of systematic uncertainty on the $\Jpsi$ yields in multiplicity classes are considered:
\begin{enumerate*}[label=(\roman*)]
  \item the signal extraction, 
  \item the normalisation, 
  \item the effect of resolution and pile-up, 
  \item the event-by-event $\Ntr$ to $\Ntrc$ correction, and
  \item the event selection efficiency of the NSD event class.
\end{enumerate*}
For the measurement of the yields in each multiplicity interval normalized to the event average, the systematic uncertainties are estimated directly for this ratio. 
Details on the signal extraction uncertainty were addressed in the previous section. The values are estimated by varying the signal and background shapes of the fit function, as well as by varying the invariant-mass range of the fit. The systematic uncertainty is computed as the root-mean-square of the uncertainties on the ratio for each of these fits, ranging between 0.8--2.3\% (0.5--1.9\%) at forward (backward) rapidity, being larger at large multiplicities where the number of events is smaller. 
The normalisation factor of the dimuon triggered to MB events $\Fnorm$ is studied using three alternative methods~\cite{Acharya:2018kxc}. 
The first method evaluates the probability of a coincidence of a dimuon- and a MB-triggered event in a MB-triggered data set.
The second method exploits the higher probability of occurrence of a  single-muon trigger by looking at the product of the probability of coincidence of a single-muon- and MB-triggered event and of the probability of finding a dimuon event in the single-muon triggered data. 
The third method is based on information from the trigger scalers. 
The run-by-run spread of the $\Fnorm/\Fnorm^i$ values, ratio of the normalisation values in the integrated and specific multiplicity intervals, computed for these three methods determines a 2.5\% systematic uncertainty, independent of multiplicity. 
The effect of the method of choice for the event-by-event correction from $\Ntr$ to $\Ntrc$ on the $\Jpsi$ yield is also studied~\cite{Abelev:2012rz,Adam:2015ota}. 
Both the randomisation function (Poisson or binomial) and the reference normalisation of the correction are varied. The Poissonian smearing is applied when the maximum is selected as normalisation reference, while the binomial correction should be used when considering all other possible reference values (in our case the minimum). The influence of these modifications on the yield ranges from 0.1\% to 2.6\% (4.3\%) at forward (backward) rapidity, as a function of multiplicity. 
The uncertainty coming from pile up and multiplicity axis resolution is estimated as a single contribution by repeating the analysis multiple times with a different randomisation seed for the event-by-event correction, or introducing a small shift of the $\Ntrc$ intervals, or varying the pile-up rejection criteria. The uncertainty amounts to 2\%, independent of multiplicity. 
The uncertainty on the event selection efficiency for the NSD event class is estimated as in Ref.~\cite{Acharya:2018egz}. The uncertainty amounts to $1\%$ and is correlated in all multiplicity intervals. 
Table~\ref{tab:unc_yield} summarizes all contributions to the systematic uncertainty on the normalized yield.

For the $\ave{\pt}$, the effects of the uncertainty on the $\ave{\pt}$ extraction procedure and of the $\Ae$ are considered.
Similar to the yields, the signal extraction uncertainty is estimated by varying the fit function and its range. In addition, as the $S/(S+B)$ terms in Eq.~\ref{eq:meanPtFit} are fixed in the fit to the $\ave{\pt}$ invariant-mass spectrum, the influence of the statistical uncertainty on the $\Jpsi$ signal $S$ is introduced via a Gaussian smearing of $S$ (with respect to its statistical uncertainty) to prevent artificially minimising the uncertainty. It ranges from 0.2\% to 3.0\% (1.2\%) at forward (backward) rapidity, increasing with multiplicity as a consequence of the smaller number of events. 
The uncertainty on the absolute $\ave{\pt}$ also takes into account the uncertainty on: 
\begin{enumerate*}[label=(\roman*)]
  \item the MC input shapes as a function of $\pt$ and $\ycms$, ranging from $<0.1$ to $6\%$ ($<0.1$ to $11\%$) at forward (backward) rapidity, 
  \item the tracking efficiency, 1\%~\cite{Acharya:2018kxc},
  \item the trigger efficiency, 2.6\% (3.1\%)~\cite{Acharya:2018kxc}, and 
  \item the matching efficiency between the tracks in the MCH and the MTR, 1\%~\cite{Acharya:2018kxc}.
\end{enumerate*}
To evaluate the uncertainty on the  MC input, the data are divided into two multiplicity classes at the mean of the $\Ntrc$ distribution for each rapidity interval. 
For each of these bins, the $\ave{\pt}$ is estimated using a modified $\Ae$ correction, which was re-weighted to better describe the $\pt$- and $y$-dependent distributions of $\Jpsi$ in given bin. The systematic uncertainty is taken as the difference of the original $\ave{\pt}$ value, computed with the initial $\Ae$ correction, and the new $\ave{\pt}$ estimated with re-weighted correction.
The uncertainty on all the measured multiplicity intervals is extrapolated from these two values assuming that in each class  the uncertainty is proportional to the $\ave{\pt}$.
The contributions of the tracking, the trigger and their matching to the uncertainty are correlated between multiplicity intervals.
The normalized $\ave{\pt}$ values are only affected by the uncertainty on the signal extraction procedure and the MC input, which is partly correlated in multiplicity and ranges from $<0.1\%$ to $2\%$ ($<0.1\%$ to $4\%$) in the forward (backward) rapidity interval. 
Table~\ref{tab:unc_pt} summarizes all contributions to the average, $\ave{\pt}$, and normalized average, $\ave{\pt} \big/ \ave{\pt^{\rm int}}$, $\pt$ measurements. 
The correlated uncertainties are added in quadrature and quoted in the plot as a text.

\begin{table}[!htbp]
   \centering
      \caption{Sources of systematic uncertainties on the normalized yield. 
   The contributions marked with an asterisk are correlated in multiplicity. 
   \label{tab:unc_yield}
   }
   \begin{tabular}{@{} c |c|c @{}} 
	\hline
	Source  				& $2.03 < \ycms < 3.53$ &  $-4.46 < \ycms < -2.96$  \\ \hline
Signal extraction 			&	$0.8$--$2.3\%$ 	& $0.5$--$1.9 \%$ 	\\
Normalization ($F_{\rm norm}$)		 &	$2.5\%$	&	$2.5\%$  \\
Event-by-event $\Ntrc$		 &	$0.1$--$2.6\%$		& $0.1$--$4.3\%$  \\
Bin-flow and pile-up			 &	$2\%$			& 		$2\% $		 \\
Normalization to NSD		 &	$1\%^{*}$			& 		$1\%^{*}$		 \\
						 \hline
   \end{tabular}
\end{table}

\begin{table}[!htbp]
   \centering
      \caption{Systematic uncertainty sources on the average and normalized average $\pt$. 
   The values in parentheses correspond to the multiplicity-integrated uncertainties related to the signal extraction. 
   The contributions marked with an asterisk are correlated in multiplicity. 
   The uncertainty on MC input, marked with a diamond, is partially correlated in multiplicity. 
   }
   \label{tab:unc_pt}
   \begin{tabular}{@{} c |cc|cc @{}} 
	\hline
	  & \multicolumn{2}{c|}{$2.03 < \ycms < 3.53$} &  \multicolumn{2}{c}{$-4.46 < \ycms < -2.96$}  \\
	Source 
		& 		$\ave{\pt}$ & $\ave{\pt} \big/ \ave{\pt^{\rm int}}$ 	
		&  		$\ave{\pt}$ & $\ave{\pt} \big/ \ave{\pt^{\rm int}}$	 \\ \hline
Signal extraction &	 	0.2--3.0\% (0.2\%)		&	0.3--3.0\%	
				& 	0.2--1.2\%	 (0.2\%)	& 	0.3--1.3\%\\
Tracking efficiency		 &	 1\%*	& --	
					 & 	1\%* 	& --	\\
Trigger efficiency		 &	2.6\%*	& --	
					 & 	 3.1\%* 	& --	\\
Track--trigger matching 	 &	 1\%*	& --	
					& 	 1\%* 	& --	\\
Monte Carlo input		 &	 $<0.1 - 6\%^{\diamond}$	& $<0.1-2\%^{\diamond}$		
				 & 	 $<0.1-11\%^{\diamond} $	& $<0.1-4\%^{\diamond}$	\\			
\hline				 		 
   \end{tabular}
\end{table}

\section{Results and discussion}
\label{sec:results}
\begin{figure}[!htbp]
   \centering
   \includegraphics[width=0.75\textwidth]{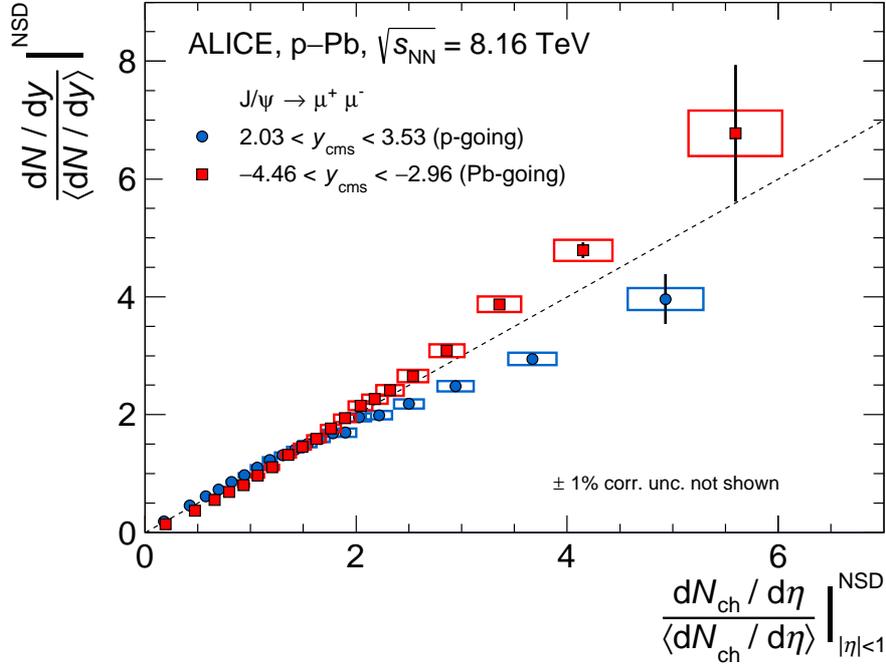} 
   \caption{
      \label{fig:JpsiYield}
      Normalized yield of inclusive $\Jpsi$, at forward and backward rapidities, as a function of the normalized charged-particle pseudorapidity density, measured at midrapidity, in \pPb collisions at \eightnnNS. 
      The vertical bars represent the statistical uncertainties. 
      The vertical and horizontal widths of the boxes represent the respective systematic uncertainties for the $\Jpsi$ yields and the multiplicities.
      The dashed line indicates the one-to-one correlation, to guide the eye.
      }
\end{figure}

The normalized $\Jpsi$ yield, at forward and backward rapidities, is presented in Fig.~\ref{fig:JpsiYield} as a function of the normalized charged-particle pseudorapidity density, measured at midrapidity ($|\eta|<1$). The normalized yield increases with increasing multiplicity in both rapidity intervals. The yield at backward rapidity grows faster than the one at forward rapidity, reaching values above those expected from a linear (with slope unity) increase at large multiplicities. 
On the other hand, at forward rapidity the values 
show a slower-than-linear increase
 at large multiplicities. 
The forward and backward rapidity yields cross a linear increase estimate (and each other) at around 1.5 times the average multiplicity. 
The underlying mechanism remains unclear. 
The forward (p-going) rapidity region probes the Pb-nucleus low Bjorken-$x$ regime ($x_{\rm Pb} \sim 10^{-5}$ in a naive 2-body calculation for $\pt=0$), whereas the backward (Pb-going) rapidity is sensitive to the intermediate-to-large values ($x_{\rm Pb} \sim 10^{-2}$). 
The observed suppression of the $\pt$- and multiplicity-integrated $\Jpsi$ yield at forward rapidity, with respect to \pp collisions, is described by different cold nuclear matter models considering the probed shadowing/saturation domain~\cite{Acharya:2018kxc}. The centrality-differential measurements at \fivenn~\cite{Adam:2015jsa} of the nuclear modification factor, $\ave\pt$ and $\ave{\pt^2}$, corresponding to relative multiplicities of at most 2.5 times the average one, can also be described by these models. 
The contribution from beauty-quark decays to the inclusive $\Jpsi$ yield amounts to $\sim10\%$~\cite{Acharya:2018yud}. It is not expected to affect significantly these results, since a similar trend was observed for prompt and non-prompt $\Jpsi$ as a function of the charged-particle pseudorapidity density in \pp~collisions~\cite{Adam:2015ota}. Moreover, the autocorrelations influence is negligible in this analysis due to the large rapidity gap between the measurement of the charged-particle multiplicity and the $\Jpsi$ yield~\cite{Weber:2018ddv}.

\begin{figure}[!htbp]
   \centering
      \includegraphics[width=0.75\textwidth]{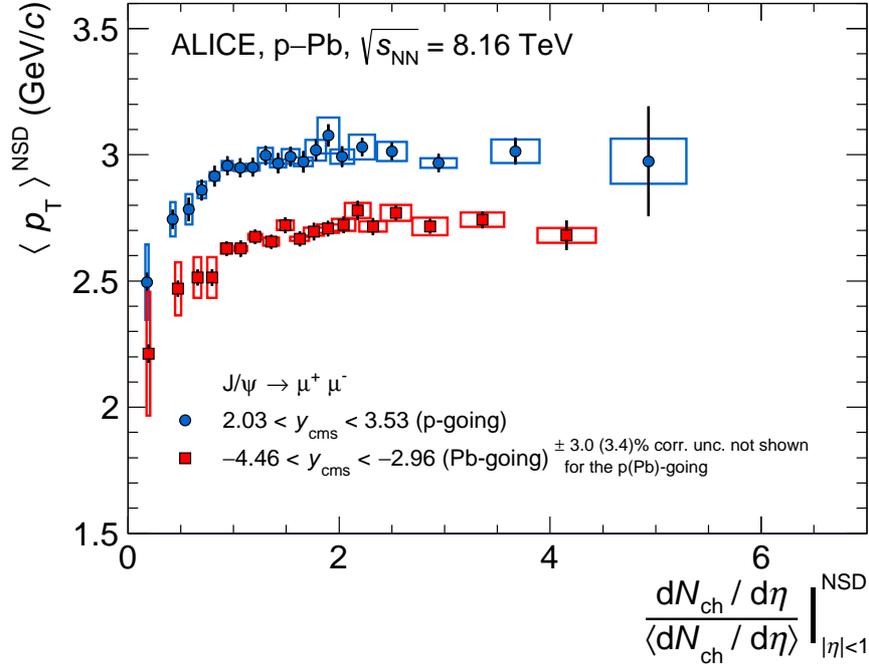}
   \caption{
      \label{fig:JpsiPt}
      Average transverse momentum of inclusive $\Jpsi$ at forward and backward rapidities as a function of the normalized charged-particle pseudorapidity density, measured at midrapidity, in \pPb collisions at \eightnnNS. 
      The vertical bars represent the statistical uncertainties, the boxes the systematic ones. 
      }
\end{figure}

Figure~\ref{fig:JpsiPt} presents $\meanpt$ as a function of the relative charged-particle pseudorapidity density, in \pPb collisions at \eightnnNS. 
The measured $\meanpt$ is systematically smaller at backward than at forward rapidity. This is also true for the multiplicity-integrated value, which is consistent with the observed decrease of $\meanpt$ with increasing $\abs{\ycms}$ in \pp~collisions~\cite{Aaij:2011jh}. 
The $\meanpt$ increases steadily for multiplicities below the average, and saturates above the average multiplicity. 
Two naive scenarios are typically considered to explain high-multiplicity events: the incoherent superposition of multiple parton--parton collisions, or single parton interactions with higher energy transfer. One would expect the latter to be characterized by a higher $\meanpt$ of the produced $\Jpsi$. 
Reality is probably somewhere in between these two simplified scenarios. 
The simultaneous increase of the \pPb~yield together with the saturation of $\meanpt$ at large multiplicities may point to $\Jpsi$ production from an incoherent superposition of parton--parton collisions. 

\begin{figure}[!htbp]
   \centering
      \includegraphics[width=0.495\textwidth]{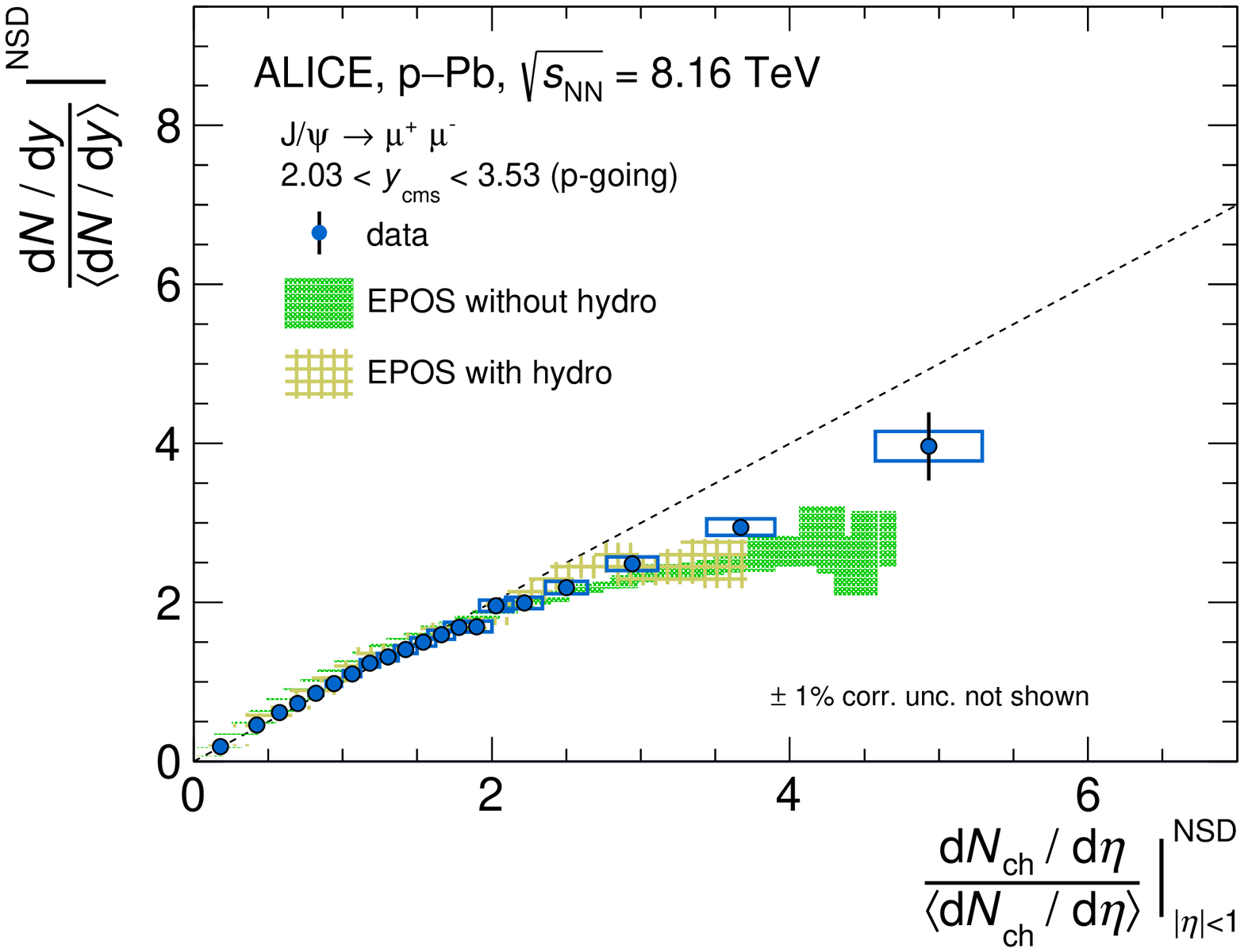} 
      \includegraphics[width=0.495\textwidth]{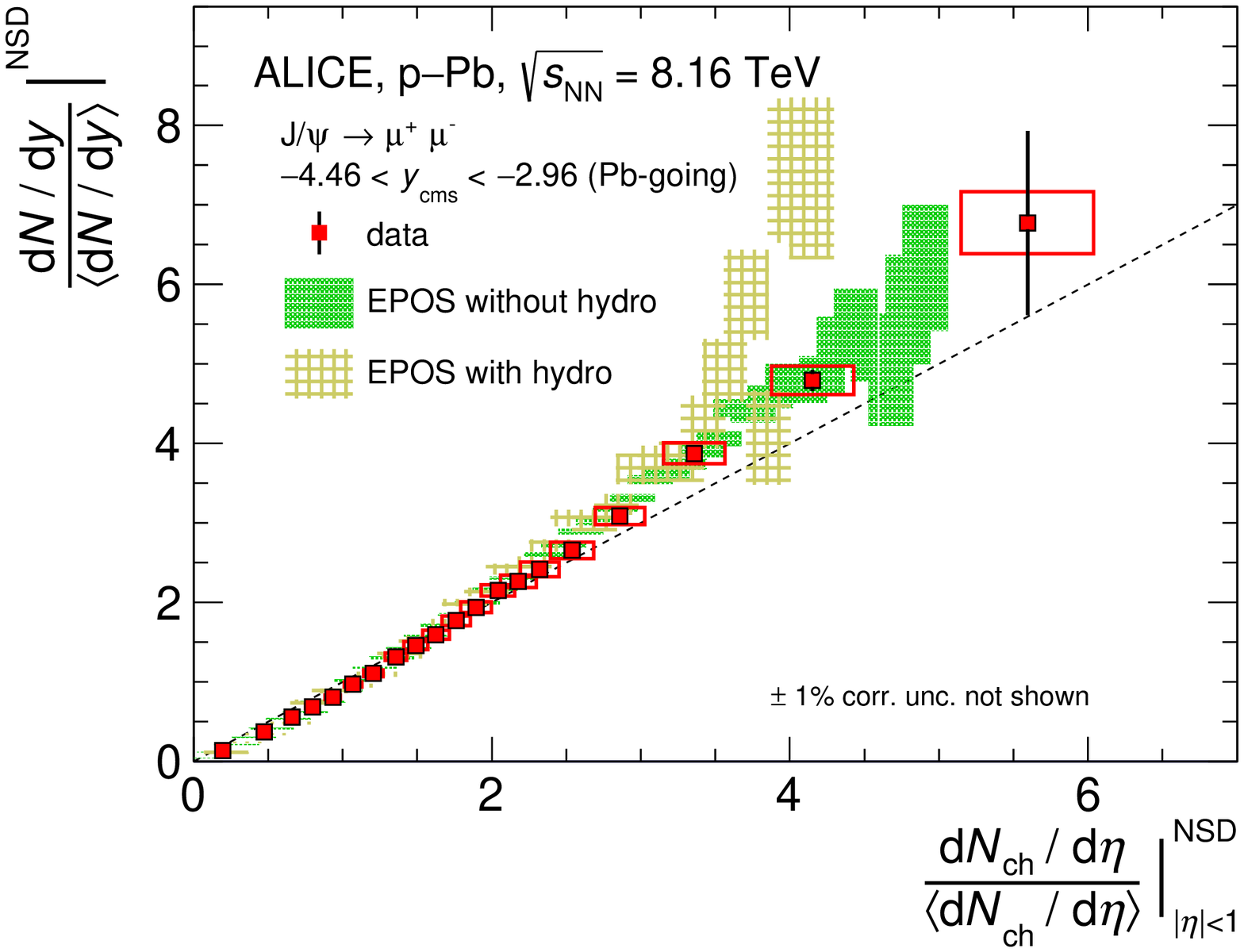} 
   \caption{
      \label{fig:JpsiYieldVsEPOS}
      Normalized yield of inclusive $\Jpsi$ as a function of the normalized charged-particle pseudorapidity density, measured at midrapidity, in \pPb collisions at \eightnnNS~compared with EPOS~3~\cite{Drescher:2000ha,Werner:2013tya} calculations. 
      Left (right) panel presents the measurement at forward (backward) rapidity.
      The vertical bars represent the statistical uncertainties, the boxes the systematic ones. 
      The dashed line indicates the one-to-one correlation, to guide the eye. 
      The shaded areas represent the statistical uncertainties on the EPOS~3 calculations. 
      }
\end{figure}

The measured yield in p--Pb collisions can be described with the EPOS~3 event generator~\cite{Drescher:2000ha,Werner:2013tya} (see Fig.~\ref{fig:JpsiYieldVsEPOS}) based on a combination of Gribov-Regge theory and pQCD: where the individual scatterings are identified with parton ladders emerging as flux tubes, the existence of multiple nucleon--nucleon collisions in \pPb~collisions is accounted for, the initial conditions of the collision are modified due to CNM effects including parton saturation, and slow string segments (far from the surface) can be further mapped to fluid dynamic fields using a core-corona description. 
The $\Jpsi$ bound-state formation in EPOS~3 assumes a color-evaporation approach, i.e.\ it is associated to a charm quark--anti-quark pair in a given mass range.
The influence of the 3D+1 viscous hydrodynamic evolution of the bulk (starting from flux tube initial conditions) in the EPOS~3 calculation is small (see Fig.~\ref{fig:JpsiYieldVsEPOS}). However the number of simulated events at large multiplicities is limited and does not allow us to elucidate possible hydrodynamic effects. 
EPOS~3~description of the measurement suggests $\Jpsi$ production from an incoherent superposition of parton--parton collisions. 

\begin{figure}[!htbp]
   \centering
      \includegraphics[width=0.6\textwidth]{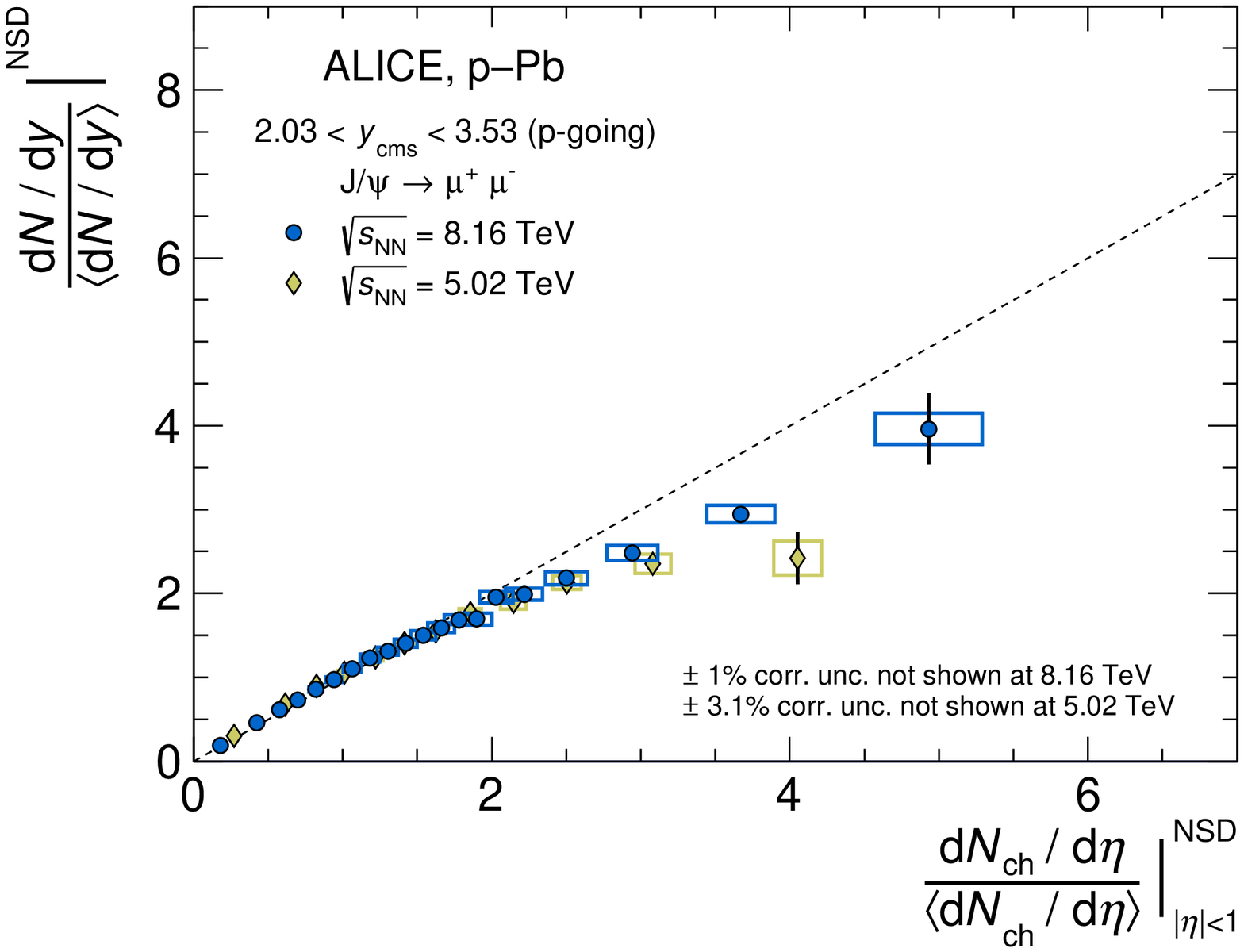} 
      \includegraphics[width=0.6\textwidth]{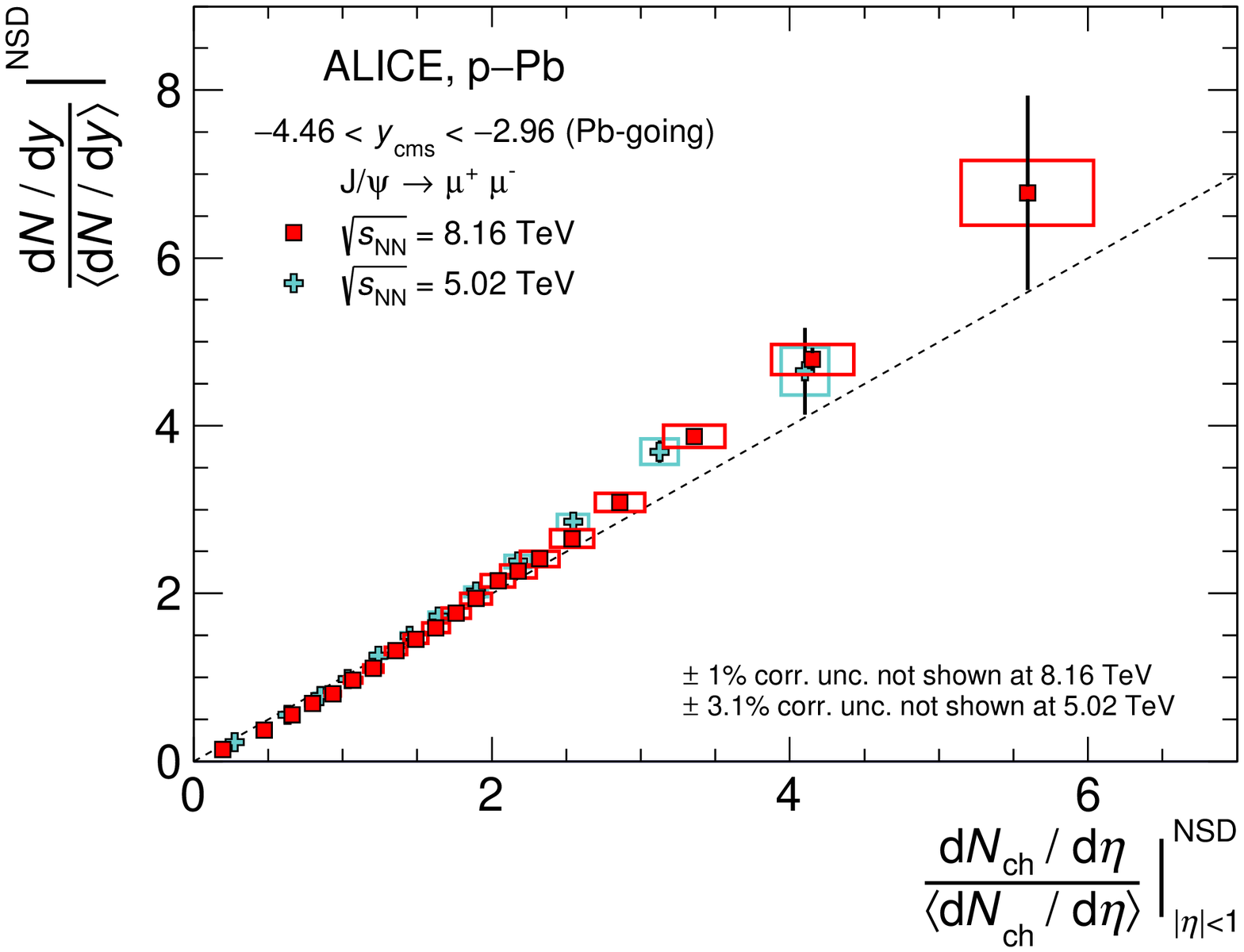} 
   \caption{
      \label{fig:JpsiYieldVsEn}
      Normalized yield of inclusive $\Jpsi$ as a function of the normalized charged-particle pseudorapidity density, measured at midrapidity, in \pPb collisions at \eightnnNS~and \fivenn~\cite{Adamova:2017uhu}. 
      The top (bottom) panel presents the measurement at forward (backward) rapidity.
      The vertical bars represent the statistical uncertainties, the boxes the systematic ones. 
      The dashed line indicates the one-to-one correlation, to guide the eye.
      }
\end{figure}

\begin{figure}[!htbp]
   \centering
      \includegraphics[width=0.6\textwidth]{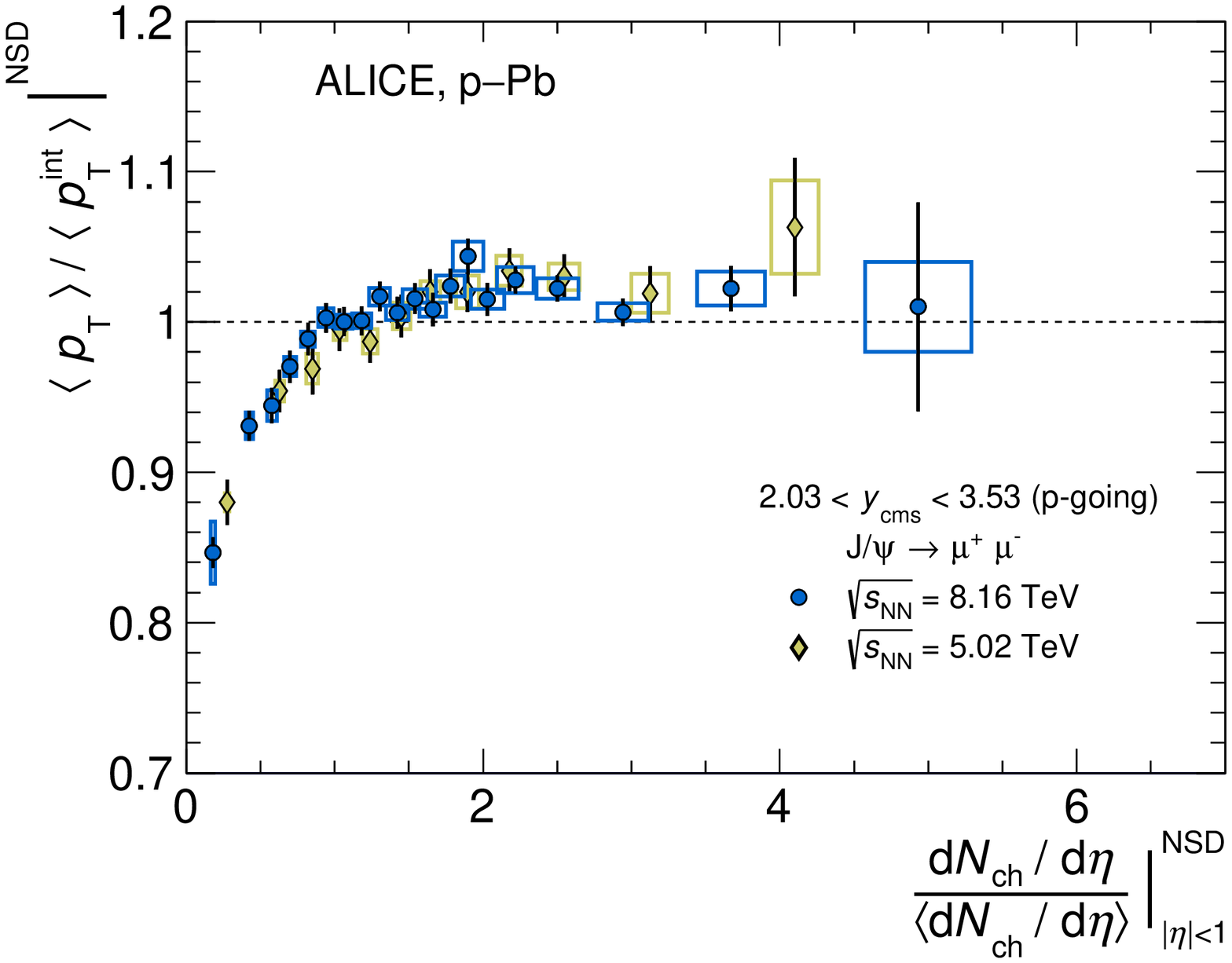} 
      \includegraphics[width=0.6\textwidth]{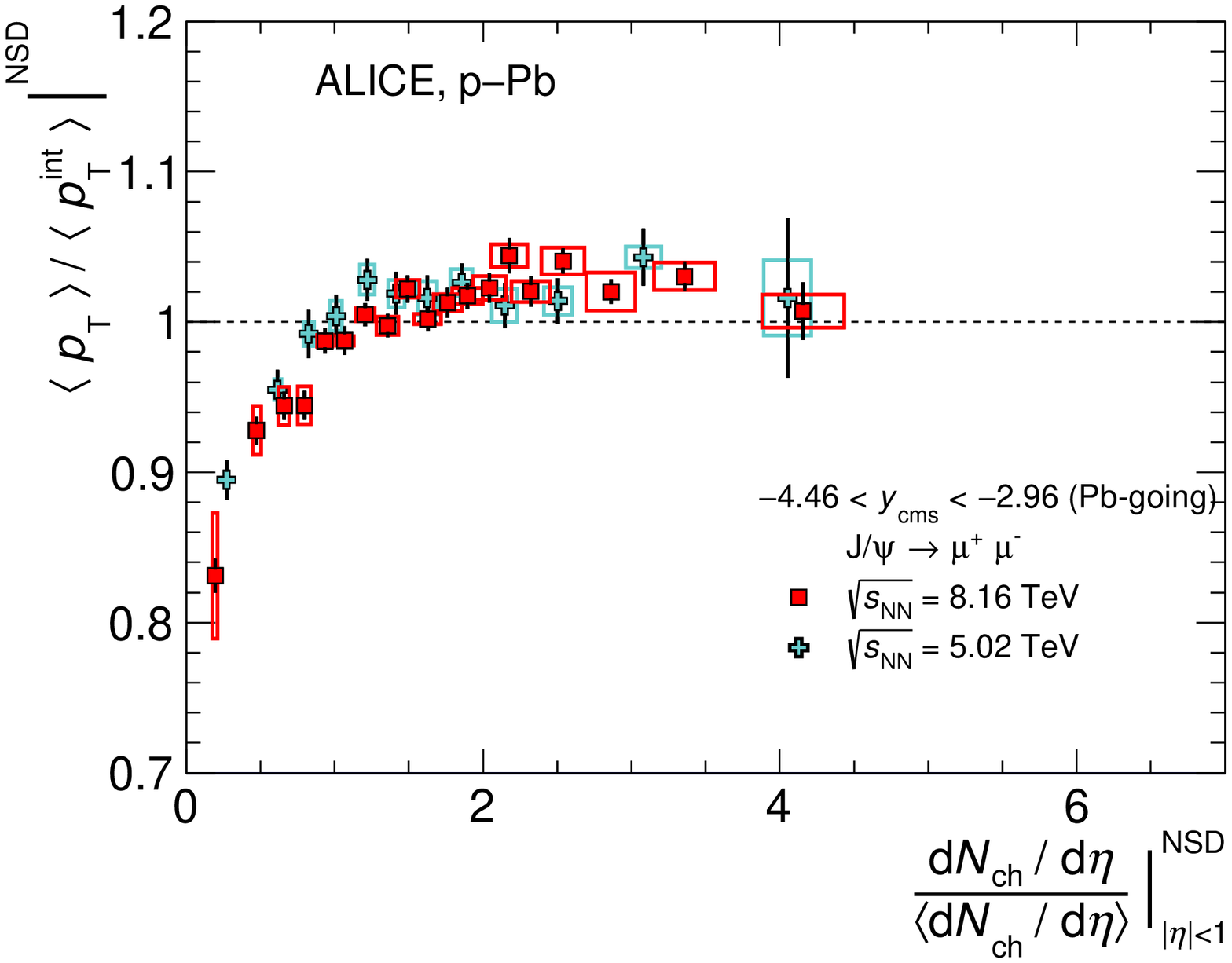} 
   \caption{
      \label{fig:JpsiPtVsEn}
       Normalized average transverse momentum of inclusive $\Jpsi$ as a function of the normalized charged-particle pseudorapidity density, measured at midrapidity, in \pPb collisions at \eightnnNS~and \fivenn~\cite{Adamova:2017uhu}. 
      Top (bottom) panel presents the measurement at forward (backward) rapidity.
      The vertical bars represent the statistical uncertainties, the boxes the systematic ones. 
      }
\end{figure}

The normalized $\Jpsi$ yield and $\meanpt$ are compared with the results in \pPb~collisions at \fivenn \cite{Adamova:2017uhu} in Fig.~\ref{fig:JpsiYieldVsEn} and \ref{fig:JpsiPtVsEn}, respectively. 
The measurements are in remarkable agreement, within the uncertainties, at both energies and rapidities.
These results extend the probed charged-particle pseudorapidity density interval, both at low and high multiplicity, examining events of up to almost six times the average value.
The more precise \eightnnNS~data evidence a continuous increase of the normalized yield with multiplicity up to the largest multiplicities attained. 
The similarities at \eightnnNS and \fivenn~suggest a common origin of the multiplicity trend, with a mechanism whose effect varies with rapidity, but might have a small dependence on the collision energy, in the explored interval. This is consistent with the large variation of the probed $x_{\rm Pb}$ with rapidity and its relative slow evolution on the collision energy (typically a factor of 2 in the simplified 2-body picture).

\begin{figure}[!htbp]
   \centering
   \includegraphics[width=0.6\textwidth]{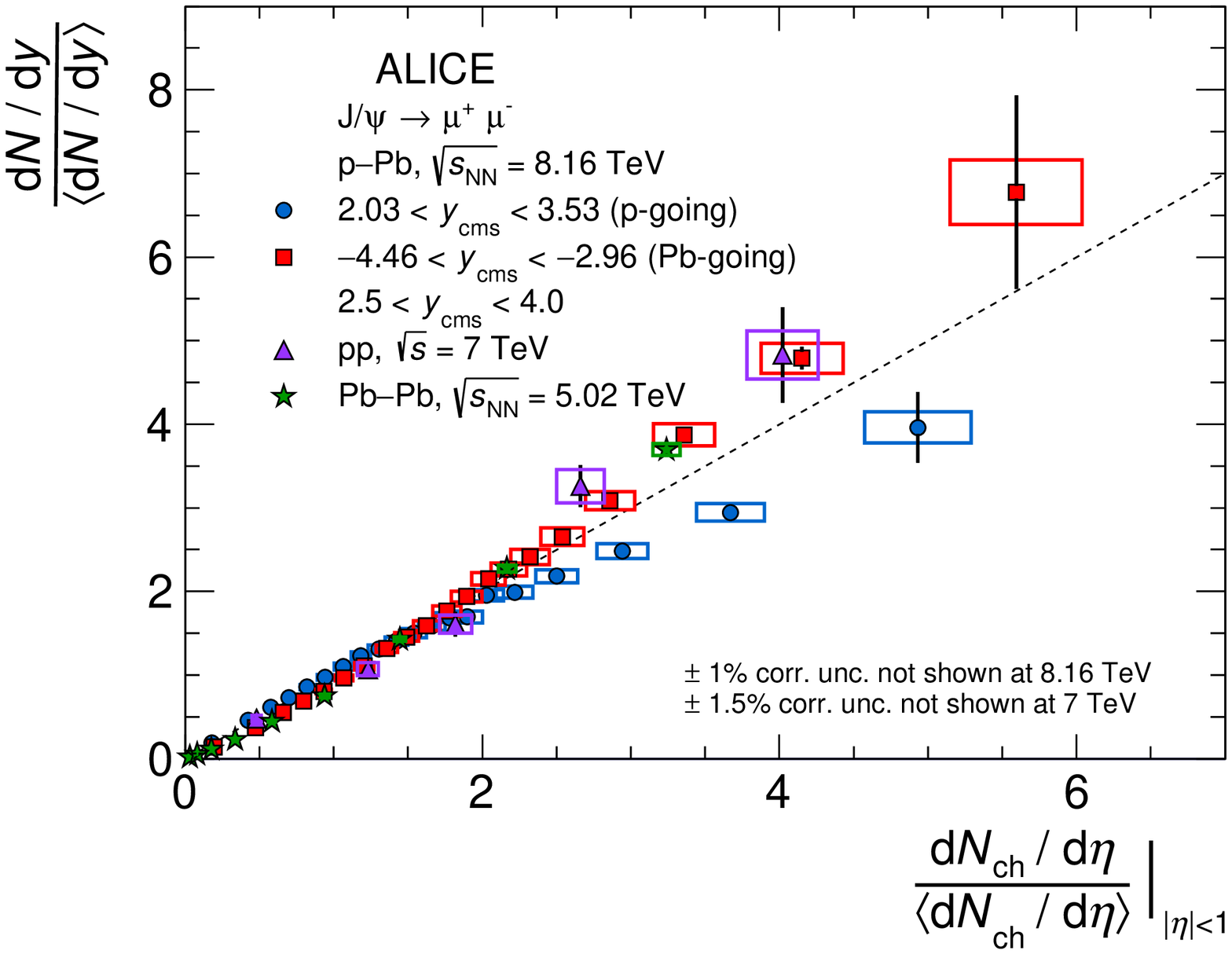} 
   \includegraphics[width=0.6\textwidth]{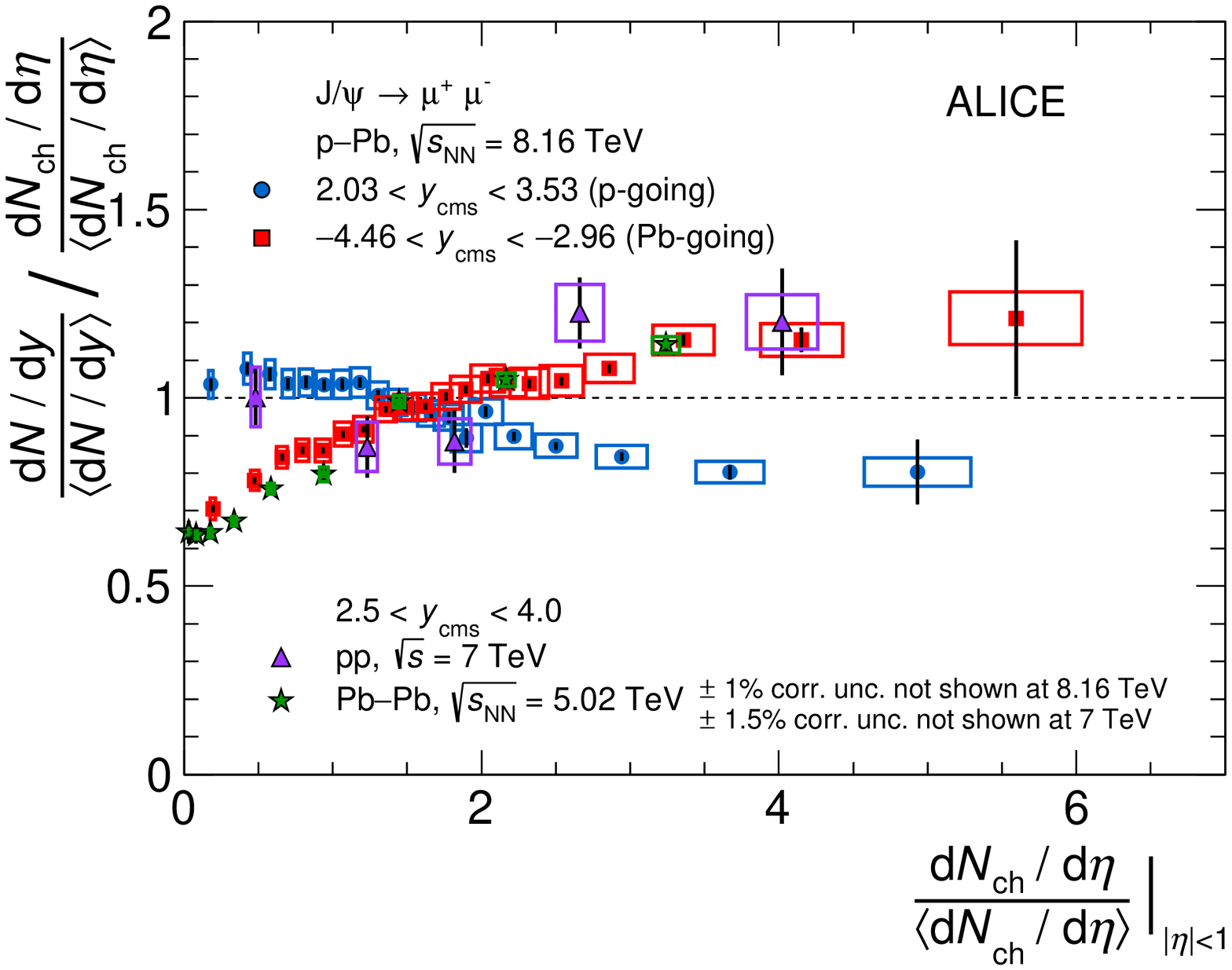} 
   \caption{
      \label{fig:JpsiYieldVsSys}
      Top: Normalized yield of inclusive $\Jpsi$ as a function of the normalized charged-particle pseudorapidity density, measured at midrapidity, in various collision systems. 
      Bottom: Ratio of the normalized yields to the corresponding normalized charged-particle pseudorapidity density.  
      The pp results are normalized to INEL collisions~\cite{Abelev:2012rz}, whereas \pPb~ones refer to the NSD event class; all for $\pt>0$. 
      The Pb--Pb data points include $\Jpsi$ with $0.3<\pt<12~{\rm GeV}/c$ to reduce the low-$\pt$ contribution from photoproduction, which is significant only in more peripheral collisions~\cite{Adam:2015gba, Acharya:2019iur,Acharya:2019vlb}.
      The vertical bars represent the statistical uncertainties, the boxes the systematic ones. 
      The dashed line indicates the one-to-one correlation, to guide the eye.
      }
\end{figure}

Figure~\ref{fig:JpsiYieldVsSys} presents a comparison of the normalized $\Jpsi$ p--Pb yields with results from pp collisions at $\sqrt{s_{\rm NN}} = 7$~TeV~\cite{Abelev:2012rz} $(2.5<\ycms<4.0)$ and Pb--Pb collisions at $\sqrt{s_{\rm NN}} = 5.02$~TeV~\cite{Acharya:2019iur} $(2.5<\ycms<4.0)$. 
The corresponding $\ave{\dndeta}$ in $|\eta|<1$ for those measurements is $6.01 \pm 0.01{\rm \, (stat.)} ^{+0.20}_{-0.12}  {\rm \, (syst.)}$~\cite{Abelev:2012rz} and $544.7 \pm 0.2 {\rm \, (stat.)} \pm 7.3 {\rm \, (syst.)}$ for 0--90\% centrality~\cite{Adam:2015ptt}, respectively. 
The ratio of the yields over the corresponding charged-particle multiplicity is also shown in Fig.~\ref{fig:JpsiYieldVsSys}. 
The trend exhibited by the pp data is similar to the one observed in the backward (Pb-going) direction. 
It should be noted that the pp results are normalized to the inelastic `INEL' event class, whereas the \pPb~measurements are normalized to the non-single-diffractive `NSD' one. In \pPb~collisions these two event classes mostly overlap when comparing MB results~\cite{Acharya:2018egz}. 
The Pb--Pb data also show a faster-than-linear increase with the normalized charged-particle pseudorapidity density. They are compatible within uncertainties with the p--Pb backward rapidity result in the restricted multiplicity interval of the measurement. 
Whereas the pp and \pPb~data include $\Jpsi$ with $\pt>0$, the Pb--Pb data points include $\Jpsi$ with $0.3<\pt<12~{\rm GeV}/c$ to reduce the low-$\pt$ contribution from photoproduction, which is significant only in more peripheral collisions~\cite{Adam:2015gba, Acharya:2019iur,Acharya:2019vlb}.
The overall increase of multiple parton--parton collisions with the colliding system (from pp up to \PbPb~collisions) described in~\cite{dEnterria:2017yhd} is expected to cancel in the relative quantities reported in this publication, sensitive to the relative evolution with charged-particle density in a given colliding system. 
Model calculations are needed to interpret the similarities of \pp, \pPb, and \PbPb~normalized $\Jpsi$ yields at large rapidity as a function of the normalized charged-particle pseudorapidity density at midrapidity.

\section{Conclusions}
\label{sec:conclusions}
The production of inclusive $\Jpsi$ at large rapidities in \pPb collisions at \eightnnNS~is reported as a function of the charged-particle pseudorapidity density at midrapidity. 
The normalized $\Jpsi$ yield shows an increase with increasing normalised charged-particle pseudorapidity density. 
The yield at backward rapidity grows faster than the forward rapidity one, reaching values above those of the linear (with slope unity) increase estimate at large normalised multiplicity, whereas the values at forward rapidity show a slower-than-linear increase. 
The trends of the normalised yield are reproduced by the EPOS~3~\cite{Drescher:2000ha,Werner:2013tya} event generator. 
The $\meanpt$ is smaller at backward than at forward rapidity, consistent with the expected softening of the spectra  with increasing $\abs{\ycms}$.
The $\meanpt$ increases steadily for multiplicities below the average, and saturates above the average multiplicity. 
The simultaneous increase of the yield together with the saturation of $\meanpt$ may point to $\Jpsi$ production from an incoherent superposition of parton--parton collisions. 
These measurements show trends compatible with those observed at \fivenn~\cite{Adamova:2017uhu} in \pPb~collisions, but in this work an improved precision and extended multiplicity coverage were reached. 
The similarities suggest a common origin, with a mechanism whose effect varies with rapidity, but with only a small dependence (if any) on the collision energy.

\newpage

\newenvironment{acknowledgement}{\relax}{\relax}
\begin{acknowledgement}
\section*{Acknowledgements}

The ALICE Collaboration would like to thank all its engineers and technicians for their invaluable contributions to the construction of the experiment and the CERN accelerator teams for the outstanding performance of the LHC complex.
The ALICE Collaboration gratefully acknowledges the resources and support provided by all Grid centres and the Worldwide LHC Computing Grid (WLCG) collaboration.
The ALICE Collaboration acknowledges the following funding agencies for their support in building and running the ALICE detector:
A. I. Alikhanyan National Science Laboratory (Yerevan Physics Institute) Foundation (ANSL), State Committee of Science and World Federation of Scientists (WFS), Armenia;
Austrian Academy of Sciences, Austrian Science Fund (FWF): [M 2467-N36] and Nationalstiftung f\"{u}r Forschung, Technologie und Entwicklung, Austria;
Ministry of Communications and High Technologies, National Nuclear Research Center, Azerbaijan;
Conselho Nacional de Desenvolvimento Cient\'{\i}fico e Tecnol\'{o}gico (CNPq), Financiadora de Estudos e Projetos (Finep), Funda\c{c}\~{a}o de Amparo \`{a} Pesquisa do Estado de S\~{a}o Paulo (FAPESP) and Universidade Federal do Rio Grande do Sul (UFRGS), Brazil;
Ministry of Education of China (MOEC) , Ministry of Science \& Technology of China (MSTC) and National Natural Science Foundation of China (NSFC), China;
Ministry of Science and Education and Croatian Science Foundation, Croatia;
Centro de Aplicaciones Tecnol\'{o}gicas y Desarrollo Nuclear (CEADEN), Cubaenerg\'{\i}a, Cuba;
Ministry of Education, Youth and Sports of the Czech Republic, Czech Republic;
The Danish Council for Independent Research | Natural Sciences, the VILLUM FONDEN and Danish National Research Foundation (DNRF), Denmark;
Helsinki Institute of Physics (HIP), Finland;
Commissariat \`{a} l'Energie Atomique (CEA) and Institut National de Physique Nucl\'{e}aire et de Physique des Particules (IN2P3) and Centre National de la Recherche Scientifique (CNRS), France;
Bundesministerium f\"{u}r Bildung und Forschung (BMBF) and GSI Helmholtzzentrum f\"{u}r Schwerionenforschung GmbH, Germany;
General Secretariat for Research and Technology, Ministry of Education, Research and Religions, Greece;
National Research, Development and Innovation Office, Hungary;
Department of Atomic Energy Government of India (DAE), Department of Science and Technology, Government of India (DST), University Grants Commission, Government of India (UGC) and Council of Scientific and Industrial Research (CSIR), India;
Indonesian Institute of Science, Indonesia;
Centro Fermi - Museo Storico della Fisica e Centro Studi e Ricerche Enrico Fermi and Istituto Nazionale di Fisica Nucleare (INFN), Italy;
Institute for Innovative Science and Technology , Nagasaki Institute of Applied Science (IIST), Japanese Ministry of Education, Culture, Sports, Science and Technology (MEXT) and Japan Society for the Promotion of Science (JSPS) KAKENHI, Japan;
Consejo Nacional de Ciencia (CONACYT) y Tecnolog\'{i}a, through Fondo de Cooperaci\'{o}n Internacional en Ciencia y Tecnolog\'{i}a (FONCICYT) and Direcci\'{o}n General de Asuntos del Personal Academico (DGAPA), Mexico;
Nederlandse Organisatie voor Wetenschappelijk Onderzoek (NWO), Netherlands;
The Research Council of Norway, Norway;
Commission on Science and Technology for Sustainable Development in the South (COMSATS), Pakistan;
Pontificia Universidad Cat\'{o}lica del Per\'{u}, Peru;
Ministry of Science and Higher Education, National Science Centre and WUT ID-UB, Poland;
Korea Institute of Science and Technology Information and National Research Foundation of Korea (NRF), Republic of Korea;
Ministry of Education and Scientific Research, Institute of Atomic Physics and Ministry of Research and Innovation and Institute of Atomic Physics, Romania;
Joint Institute for Nuclear Research (JINR), Ministry of Education and Science of the Russian Federation, National Research Centre Kurchatov Institute, Russian Science Foundation and Russian Foundation for Basic Research, Russia;
Ministry of Education, Science, Research and Sport of the Slovak Republic, Slovakia;
National Research Foundation of South Africa, South Africa;
Swedish Research Council (VR) and Knut \& Alice Wallenberg Foundation (KAW), Sweden;
European Organization for Nuclear Research, Switzerland;
Suranaree University of Technology (SUT), National Science and Technology Development Agency (NSDTA) and Office of the Higher Education Commission under NRU project of Thailand, Thailand;
Turkish Atomic Energy Agency (TAEK), Turkey;
National Academy of  Sciences of Ukraine, Ukraine;
Science and Technology Facilities Council (STFC), United Kingdom;
National Science Foundation of the United States of America (NSF) and United States Department of Energy, Office of Nuclear Physics (DOE NP), United States of America.
\end{acknowledgement}

\bibliographystyle{utphys}   
\bibliography{bibliography}

\newpage
\appendix

%
%

\section{The ALICE Collaboration}
\label{app:collab}

\begingroup
\small
\begin{flushleft}
S.~Acharya\Irefn{org141}\And 
D.~Adamov\'{a}\Irefn{org95}\And 
A.~Adler\Irefn{org74}\And 
J.~Adolfsson\Irefn{org81}\And 
M.M.~Aggarwal\Irefn{org100}\And 
G.~Aglieri Rinella\Irefn{org34}\And 
M.~Agnello\Irefn{org30}\And 
N.~Agrawal\Irefn{org10}\textsuperscript{,}\Irefn{org54}\And 
Z.~Ahammed\Irefn{org141}\And 
S.~Ahmad\Irefn{org16}
% ...
\And
\renewcommand\labelenumi{\textsuperscript{\theenumi}~}

\section*{Affiliation notes}
\renewcommand\theenumi{\roman{enumi}}
\begin{Authlist}
\item \Adef{org*}Deceased
\item \Adef{orgI}Italian National Agency for New Technologies, Energy and Sustainable Economic Development (ENEA), Bologna, Italy
\item \Adef{orgII}Dipartimento DET del Politecnico di Torino, Turin, Italy
\item \Adef{orgIII}M.V. Lomonosov Moscow State University, D.V. Skobeltsyn Institute of Nuclear, Physics, Moscow, Russia
\item \Adef{orgIV}Department of Applied Physics, Aligarh Muslim University, Aligarh, India
\item \Adef{orgV}Institute of Theoretical Physics, University of Wroclaw, Poland
\end{Authlist}

\section*{Collaboration Institutes}
\renewcommand\theenumi{\arabic{enumi}~}
\begin{Authlist}
\item \Idef{org1}A.I. Alikhanyan National Science Laboratory (Yerevan Physics Institute) Foundation, Yerevan, Armenia
\item \Idef{org2}Bogolyubov Institute for Theoretical Physics, National Academy of Sciences of Ukraine, Kiev, Ukraine
\item \Idef{org3}Bose Institute, Department of Physics  and Centre for Astroparticle Physics and Space Science (CAPSS), Kolkata, India
\item \Idef{org4}Budker Institute for Nuclear Physics, Novosibirsk, Russia
\item \Idef{org5}California Polytechnic State University, San Luis Obispo, California, United States
\item \Idef{org6}Central China Normal University, Wuhan, China
\item \Idef{org7}Centre de Calcul de l'IN2P3, Villeurbanne, Lyon, France
\item \Idef{org8}Centro de Aplicaciones Tecnol\'{o}gicas y Desarrollo Nuclear (CEADEN), Havana, Cuba
\item \Idef{org9}Centro de Investigaci\'{o}n y de Estudios Avanzados (CINVESTAV), Mexico City and M\'{e}rida, Mexico
\item \Idef{org10}Centro Fermi - Museo Storico della Fisica e Centro Studi e Ricerche ``Enrico Fermi', Rome, Italy
\item \Idef{org11}Chicago State University, Chicago, Illinois, United States
\item \Idef{org12}China Institute of Atomic Energy, Beijing, China
\item \Idef{org13}Comenius University Bratislava, Faculty of Mathematics, Physics and Informatics, Bratislava, Slovakia
\item \Idef{org14}COMSATS University Islamabad, Islamabad, Pakistan
\item \Idef{org15}Creighton University, Omaha, Nebraska, United States
\item \Idef{org16}Department of Physics, Aligarh Muslim University, Aligarh, India
\item \Idef{org17}Department of Physics, Pusan National University, Pusan, Republic of Korea
\item \Idef{org18}Department of Physics, Sejong University, Seoul, Republic of Korea
\item \Idef{org19}Department of Physics, University of California, Berkeley, California, United States
\item \Idef{org20}Department of Physics, University of Oslo, Oslo, Norway
\item \Idef{org21}Department of Physics and Technology, University of Bergen, Bergen, Norway
\item \Idef{org22}Dipartimento di Fisica dell'Universit\`{a} 'La Sapienza' and Sezione INFN, Rome, Italy
\item \Idef{org23}Dipartimento di Fisica dell'Universit\`{a} and Sezione INFN, Cagliari, Italy
\item \Idef{org24}Dipartimento di Fisica dell'Universit\`{a} and Sezione INFN, Trieste, Italy
\item \Idef{org25}Dipartimento di Fisica dell'Universit\`{a} and Sezione INFN, Turin, Italy
\item \Idef{org26}Dipartimento di Fisica e Astronomia dell'Universit\`{a} and Sezione INFN, Bologna, Italy
\item \Idef{org27}Dipartimento di Fisica e Astronomia dell'Universit\`{a} and Sezione INFN, Catania, Italy
\item \Idef{org28}Dipartimento di Fisica e Astronomia dell'Universit\`{a} and Sezione INFN, Padova, Italy
\item \Idef{org29}Dipartimento di Fisica `E.R.~Caianiello' dell'Universit\`{a} and Gruppo Collegato INFN, Salerno, Italy
\item \Idef{org30}Dipartimento DISAT del Politecnico and Sezione INFN, Turin, Italy
\item \Idef{org31}Dipartimento di Scienze e Innovazione Tecnologica dell'Universit\`{a} del Piemonte Orientale and INFN Sezione di Torino, Alessandria, Italy
\item \Idef{org32}Dipartimento di Scienze MIFT, Universit\`{a} di Messina, Messina, Italy
\item \Idef{org33}Dipartimento Interateneo di Fisica `M.~Merlin' and Sezione INFN, Bari, Italy
\item \Idef{org34}European Organization for Nuclear Research (CERN), Geneva, Switzerland
\item \Idef{org35}Faculty of Electrical Engineering, Mechanical Engineering and Naval Architecture, University of Split, Split, Croatia
\item \Idef{org36}Faculty of Engineering and Science, Western Norway University of Applied Sciences, Bergen, Norway
\item \Idef{org37}Faculty of Nuclear Sciences and Physical Engineering, Czech Technical University in Prague, Prague, Czech Republic
\item \Idef{org38}Faculty of Science, P.J.~\v{S}af\'{a}rik University, Ko\v{s}ice, Slovakia
\item \Idef{org39}Frankfurt Institute for Advanced Studies, Johann Wolfgang Goethe-Universit\"{a}t Frankfurt, Frankfurt, Germany
\item \Idef{org40}Fudan University, Shanghai, China
\item \Idef{org41}Gangneung-Wonju National University, Gangneung, Republic of Korea
\item \Idef{org42}Gauhati University, Department of Physics, Guwahati, India
\item \Idef{org43}Helmholtz-Institut f\"{u}r Strahlen- und Kernphysik, Rheinische Friedrich-Wilhelms-Universit\"{a}t Bonn, Bonn, Germany
\item \Idef{org44}Helsinki Institute of Physics (HIP), Helsinki, Finland
\item \Idef{org45}High Energy Physics Group,  Universidad Aut\'{o}noma de Puebla, Puebla, Mexico
\item \Idef{org46}Hiroshima University, Hiroshima, Japan
\item \Idef{org47}Hochschule Worms, Zentrum  f\"{u}r Technologietransfer und Telekommunikation (ZTT), Worms, Germany
\item \Idef{org48}Horia Hulubei National Institute of Physics and Nuclear Engineering, Bucharest, Romania
\item \Idef{org49}Indian Institute of Technology Bombay (IIT), Mumbai, India
\item \Idef{org50}Indian Institute of Technology Indore, Indore, India
\item \Idef{org51}Indonesian Institute of Sciences, Jakarta, Indonesia
\item \Idef{org52}INFN, Laboratori Nazionali di Frascati, Frascati, Italy
\item \Idef{org53}INFN, Sezione di Bari, Bari, Italy
\item \Idef{org54}INFN, Sezione di Bologna, Bologna, Italy
\item \Idef{org55}INFN, Sezione di Cagliari, Cagliari, Italy
\item \Idef{org56}INFN, Sezione di Catania, Catania, Italy
\item \Idef{org57}INFN, Sezione di Padova, Padova, Italy
\item \Idef{org58}INFN, Sezione di Roma, Rome, Italy
\item \Idef{org59}INFN, Sezione di Torino, Turin, Italy
\item \Idef{org60}INFN, Sezione di Trieste, Trieste, Italy
\item \Idef{org61}Inha University, Incheon, Republic of Korea
\item \Idef{org62}Institute for Nuclear Research, Academy of Sciences, Moscow, Russia
\item \Idef{org63}Institute for Subatomic Physics, Utrecht University/Nikhef, Utrecht, Netherlands
\item \Idef{org64}Institute of Experimental Physics, Slovak Academy of Sciences, Ko\v{s}ice, Slovakia
\item \Idef{org65}Institute of Physics, Homi Bhabha National Institute, Bhubaneswar, India
\item \Idef{org66}Institute of Physics of the Czech Academy of Sciences, Prague, Czech Republic
\item \Idef{org67}Institute of Space Science (ISS), Bucharest, Romania
\item \Idef{org68}Institut f\"{u}r Kernphysik, Johann Wolfgang Goethe-Universit\"{a}t Frankfurt, Frankfurt, Germany
\item \Idef{org69}Instituto de Ciencias Nucleares, Universidad Nacional Aut\'{o}noma de M\'{e}xico, Mexico City, Mexico
\item \Idef{org70}Instituto de F\'{i}sica, Universidade Federal do Rio Grande do Sul (UFRGS), Porto Alegre, Brazil
\item \Idef{org71}Instituto de F\'{\i}sica, Universidad Nacional Aut\'{o}noma de M\'{e}xico, Mexico City, Mexico
\item \Idef{org72}iThemba LABS, National Research Foundation, Somerset West, South Africa
\item \Idef{org73}Jeonbuk National University, Jeonju, Republic of Korea
\item \Idef{org74}Johann-Wolfgang-Goethe Universit\"{a}t Frankfurt Institut f\"{u}r Informatik, Fachbereich Informatik und Mathematik, Frankfurt, Germany
\item \Idef{org75}Joint Institute for Nuclear Research (JINR), Dubna, Russia
\item \Idef{org76}Korea Institute of Science and Technology Information, Daejeon, Republic of Korea
\item \Idef{org77}KTO Karatay University, Konya, Turkey
\item \Idef{org78}Laboratoire de Physique des 2 Infinis, Irène Joliot-Curie, Orsay, France
\item \Idef{org79}Laboratoire de Physique Subatomique et de Cosmologie, Universit\'{e} Grenoble-Alpes, CNRS-IN2P3, Grenoble, France
\item \Idef{org80}Lawrence Berkeley National Laboratory, Berkeley, California, United States
\item \Idef{org81}Lund University Department of Physics, Division of Particle Physics, Lund, Sweden
\item \Idef{org82}Nagasaki Institute of Applied Science, Nagasaki, Japan
\item \Idef{org83}Nara Women{'}s University (NWU), Nara, Japan
\item \Idef{org84}National and Kapodistrian University of Athens, School of Science, Department of Physics , Athens, Greece
\item \Idef{org85}National Centre for Nuclear Research, Warsaw, Poland
\item \Idef{org86}National Institute of Science Education and Research, Homi Bhabha National Institute, Jatni, India
\item \Idef{org87}National Nuclear Research Center, Baku, Azerbaijan
\item \Idef{org88}National Research Centre Kurchatov Institute, Moscow, Russia
\item \Idef{org89}Niels Bohr Institute, University of Copenhagen, Copenhagen, Denmark
\item \Idef{org90}Nikhef, National institute for subatomic physics, Amsterdam, Netherlands
\item \Idef{org91}NRC Kurchatov Institute IHEP, Protvino, Russia
\item \Idef{org92}NRC «Kurchatov Institute»  - ITEP, Moscow, Russia
\item \Idef{org93}NRNU Moscow Engineering Physics Institute, Moscow, Russia
\item \Idef{org94}Nuclear Physics Group, STFC Daresbury Laboratory, Daresbury, United Kingdom
\item \Idef{org95}Nuclear Physics Institute of the Czech Academy of Sciences, \v{R}e\v{z} u Prahy, Czech Republic
\item \Idef{org96}Oak Ridge National Laboratory, Oak Ridge, Tennessee, United States
\item \Idef{org97}Ohio State University, Columbus, Ohio, United States
\item \Idef{org98}Petersburg Nuclear Physics Institute, Gatchina, Russia
\item \Idef{org99}Physics department, Faculty of science, University of Zagreb, Zagreb, Croatia
\item \Idef{org100}Physics Department, Panjab University, Chandigarh, India
\item \Idef{org101}Physics Department, University of Jammu, Jammu, India
\item \Idef{org102}Physics Department, University of Rajasthan, Jaipur, India
\item \Idef{org103}Physikalisches Institut, Eberhard-Karls-Universit\"{a}t T\"{u}bingen, T\"{u}bingen, Germany
\item \Idef{org104}Physikalisches Institut, Ruprecht-Karls-Universit\"{a}t Heidelberg, Heidelberg, Germany
\item \Idef{org105}Physik Department, Technische Universit\"{a}t M\"{u}nchen, Munich, Germany
\item \Idef{org106}Politecnico di Bari, Bari, Italy
\item \Idef{org107}Research Division and ExtreMe Matter Institute EMMI, GSI Helmholtzzentrum f\"ur Schwerionenforschung GmbH, Darmstadt, Germany
\item \Idef{org108}Rudjer Bo\v{s}kovi\'{c} Institute, Zagreb, Croatia
\item \Idef{org109}Russian Federal Nuclear Center (VNIIEF), Sarov, Russia
\item \Idef{org110}Saha Institute of Nuclear Physics, Homi Bhabha National Institute, Kolkata, India
\item \Idef{org111}School of Physics and Astronomy, University of Birmingham, Birmingham, United Kingdom
\item \Idef{org112}Secci\'{o}n F\'{\i}sica, Departamento de Ciencias, Pontificia Universidad Cat\'{o}lica del Per\'{u}, Lima, Peru
\item \Idef{org113}St. Petersburg State University, St. Petersburg, Russia
\item \Idef{org114}Stefan Meyer Institut f\"{u}r Subatomare Physik (SMI), Vienna, Austria
\item \Idef{org115}SUBATECH, IMT Atlantique, Universit\'{e} de Nantes, CNRS-IN2P3, Nantes, France
\item \Idef{org116}Suranaree University of Technology, Nakhon Ratchasima, Thailand
\item \Idef{org117}Technical University of Ko\v{s}ice, Ko\v{s}ice, Slovakia
\item \Idef{org118}The Henryk Niewodniczanski Institute of Nuclear Physics, Polish Academy of Sciences, Cracow, Poland
\item \Idef{org119}The University of Texas at Austin, Austin, Texas, United States
\item \Idef{org120}Universidad Aut\'{o}noma de Sinaloa, Culiac\'{a}n, Mexico
\item \Idef{org121}Universidade de S\~{a}o Paulo (USP), S\~{a}o Paulo, Brazil
\item \Idef{org122}Universidade Estadual de Campinas (UNICAMP), Campinas, Brazil
\item \Idef{org123}Universidade Federal do ABC, Santo Andre, Brazil
\item \Idef{org124}University of Cape Town, Cape Town, South Africa
\item \Idef{org125}University of Houston, Houston, Texas, United States
\item \Idef{org126}University of Jyv\"{a}skyl\"{a}, Jyv\"{a}skyl\"{a}, Finland
\item \Idef{org127}University of Liverpool, Liverpool, United Kingdom
\item \Idef{org128}University of Science and Technology of China, Hefei, China
\item \Idef{org129}University of South-Eastern Norway, Tonsberg, Norway
\item \Idef{org130}University of Tennessee, Knoxville, Tennessee, United States
\item \Idef{org131}University of the Witwatersrand, Johannesburg, South Africa
\item \Idef{org132}University of Tokyo, Tokyo, Japan
\item \Idef{org133}University of Tsukuba, Tsukuba, Japan
\item \Idef{org134}Universit\'{e} Clermont Auvergne, CNRS/IN2P3, LPC, Clermont-Ferrand, France
\item \Idef{org135}Universit\'{e} de Lyon, Universit\'{e} Lyon 1, CNRS/IN2P3, IPN-Lyon, Villeurbanne, Lyon, France
\item \Idef{org136}Universit\'{e} de Strasbourg, CNRS, IPHC UMR 7178, F-67000 Strasbourg, France, Strasbourg, France
\item \Idef{org137}Universit\'{e} Paris-Saclay Centre d'Etudes de Saclay (CEA), IRFU, D\'{e}partment de Physique Nucl\'{e}aire (DPhN), Saclay, France
\item \Idef{org138}Universit\`{a} degli Studi di Foggia, Foggia, Italy
\item \Idef{org139}Universit\`{a} degli Studi di Pavia, Pavia, Italy
\item \Idef{org140}Universit\`{a} di Brescia, Brescia, Italy
\item \Idef{org141}Variable Energy Cyclotron Centre, Homi Bhabha National Institute, Kolkata, India
\item \Idef{org142}Warsaw University of Technology, Warsaw, Poland
\item \Idef{org143}Wayne State University, Detroit, Michigan, United States
\item \Idef{org144}Westf\"{a}lische Wilhelms-Universit\"{a}t M\"{u}nster, Institut f\"{u}r Kernphysik, M\"{u}nster, Germany
\item \Idef{org145}Wigner Research Centre for Physics, Budapest, Hungary
\item \Idef{org146}Yale University, New Haven, Connecticut, United States
\item \Idef{org147}Yonsei University, Seoul, Republic of Korea
\end{Authlist}
\endgroup
  
\end{document}